\documentclass[11pt]{article}
\usepackage{amssymb,amsmath,amsfonts}
\usepackage{graphicx}
\usepackage{graphics}
\usepackage{eepic,epsfig}

\@addtoreset{equation}{section}

\makeatother

\begin{document}

\title{Scalar Casimir densities and forces for parallel plates \\
in cosmic string spacetime}
\author{E. R. Bezerra de Mello$^{1}$\thanks{%
E-mail: emello@fisica.ufpb.br}, \thinspace\ A. A. Saharian$^{2}$\thanks{%
E-mail: saharian@ysu.am}, \thinspace\ S. V. Abajyan$^{3}$\thanks{%
E-mail: samvel.abajyan@mail.ru}, \vspace{0.3cm} \\
\textit{$^{1}$Departamento de F\'{\i}sica, Universidade Federal da Para\'{\i}%
ba}\\
\textit{58.059-970, Caixa Postal 5.008, Jo\~{a}o Pessoa, PB, Brazil}\vspace{%
0.3cm}\\
\textit{$^2$Department of Physics, Yerevan State University,}\\
\textit{1 Alex Manoogian Street, 0025 Yerevan, Armenia} \vspace{0.3cm}\\
\textit{$^3$Armenian State Pedagogical University,}\\
\textit{13 Khandjyan Street, 0010 Yerevan, Armenia}}
\date{}
\maketitle

\begin{abstract}
We analyze the Green function, the Casimir densities and forces associated
with a massive scalar quantum field confined between two parallel plates in
a higher dimensional cosmic string spacetime. The plates are placed
orthogonal to the string and the field obeys the Robin boundary conditions
on them. The boundary-induced contributions are explicitly extracted in the
vacuum expectation values (VEVs) of the field squared and of the
energy-momentum tensor for both the single plate and two plates geometries.
The VEV of the energy-momentum tensor, in additional to the diagonal
components, contains an off-diagonal component corresponding to the shear
stress. The latter vanishes on the plates in special cases of Dirichlet and
Neumann boundary conditions. For points outside the string core the
topological contributions in the VEVs are finite on the plates. Near the
string the VEVs are dominated by the boundary-free part, whereas at large
distances the boundary-induced contributions dominate. Due to the nonzero
off-diagonal component of the vacuum energy-momentum tensor, in addition to
the normal component, the Casimir forces have nonzero component parallel to
the boundary (shear force). Unlike the problem on the Minkowski bulk, the
normal forces acting on the separate plates, in general, do not coincide if
the corresponding Robin coefficients are different. Another difference is
that in the presence of the cosmic string the Casimir forces for Dirichlet
and Neumann boundary conditions differ. For Dirichlet boundary condition the
normal Casimir force does not depend on the curvature coupling parameter.
This is not the case for other boundary conditions. A new qualitative
feature induced by the cosmic string is the appearance of the shear stress
acting on the plates. The corresponding force is directed along the radial
coordinate and vanishes for Dirichlet and Neumann boundary conditions.
Depending on the parameters of the problem, the radial component of the
shear force can be either positive or negative.
\end{abstract}

\bigskip


\bigskip

\section{Introduction}

\label{sec2}

Among the most interesting consequences of phase transitions in gauge
theories is the formation of a variety of topological defects \cite{Vile94}.
The type of defect formed depends on the nature of symmetry breaking. In
particular, due to their important role in cosmology, the cosmic strings are
most thoroughly studied in the literature. The early interest to this class
of topological defects was motivated by the scenario of the large-scale
structure formation in the Universe where the strings seed the primordial
density perturbations. In the eighties this was the most popular alternative
to the inflationary scenario based on quantum fluctuations of fields during
the inflation. Although the further observations of the temperature
anisotropies of the cosmic microwave background radiation (CMB) excluded the
cosmic strings as the main source for the density perturbations, this type
of topological defects are still candidates for the generation of a number
of interesting effects that include the generation of gamma ray bursts,
high-energy cosmic rays and gravitational waves. Among the other observable
consequences we can mention here the gravitational lensing, the creation of
small non-Gaussianities in the CMB and some influence on the corresponding
tensor modes.

Depending on the underlying microscopic model, the cosmic strings can be
realized as nontrivial field configurations or they can be fundamental
quantum strings stretched to cosmological scales (cosmic superstrings, first
considered in \cite{Witt85}). A mechanism for the generation of the latter
type of objects with low values of the string tensions has been recently
proposed within the framework of brane inflationary models (see, for
instance, \cite{Cope11} and references therein). Defects of the cosmic
string-type appear also in a number of condensed matter systems such as
crystals, liquid crystals and quantum liquids \cite{Nels02}. Although the
specific properties of cosmic strings are model-dependent, they produce
similar gravitational effects. In the simplified model of a straight cosmic
string and at large distances from the string, compared with the core
radius, these effects generate planar angle deficit in the plane
perpendicular to the string.

The nontrivial topology of the cosmic string spacetime provides a distortion
of the spectrum for vacuum fluctuations of quantized fields. As a
consequence, the vacuum expectation values (VEVs) of physical observables
are shifted. Explicit calculations have been done for the field squared and
energy-momentum tensor in the cases of scalar, fermionic and electromagnetic
fields (see, for instance, references in \cite{Bell14}). For charged fields
and for cosmic strings carrying magnetic flux, other important
characteristics of the vacuum state, influenced by the planar angle deficit,
are the charge and current densities. The vacuum polarization induced by a
cosmic string in background of curved maximally symmetric spacetimes, namely
in de Sitter and anti-de Sitter spacetimes, has been discussed in \cite%
{Beze09dS} and \cite{Beze12adS}. For the background Schwarzschild spacetime
threaded by an infinite straight cosmic string this phenomenon is
investigated in \cite{Otte10}.

In a number of problems with cosmic strings additional boundaries are
present on which the operators of quantum fields obey prescribed boundary
conditions. Examples are the branes in brane inflationary models with cosmic
superstrings. The imposition of boundary conditions on quantum fields gives
rise additional shifts in the VEVs. This is the well known Casimir effect
(for reviews see \cite{Most97}). It has been investigated for a large number
of bulk and boundary geometries and has been confirmed experimentally with
high accuracy. For a cylindrical shell coaxial with the string, the combined
quantum effects of the topology and boundaries have been considered for
scalar \cite{Beze06sc,Nest11}, fermionic \cite{Beze08,Beze10}, and
electromagnetic fields \cite{Nest11,Brev95,Beze07}. The Casimir force for a
massless scalar fields subject to Dirichlet and Neumann boundary conditions
in the setting of the conical piston has been discussed in \cite{Fucc11}.
The Casimir densities for scalar and electromagnetic fields induced by
boundaries perpendicular to the string were considered in \cite%
{Beze11,Beze12,Beze13,Grig17}. Another type of boundary conditions arise in
models with cosmic strings compactified along the axis. The influence of
this compactification on the properties of the quantum vacuum has been
discussed in \cite{Beze12Comp}.

In the present paper we are interested in the analysis of the influence of a
cosmic string on the vacuum properties for a scalar field confined between
two parallel plates. The plates are perpendicular to the cosmic string and
on them the field operator obeys Robin boundary conditions, in general, with
different coefficients for separate plates. Motivated by possible
applications for cosmic superstrings, the problem will be considered in an
arbitrary number of spatial dimensions. The paper is organized as follows.
In the next section we present the problem formulation and the evaluation of
the heat kernel for a scalar field in the region between the plates. By
using the heat kernel method, in section \ref{sec:Green}, a representation
of the Green function is provided with explicitly extracted boundary-free
topological contribution. For points away from the plates, the
renormalization in the coincidence limit is required for that contribution
only. The VEVs of the field squared and of the energy-momentum tensor in the
presence of a single plate are investigated in section \ref{sec:VEV1pl}.
This section generalizes the results obtained in \cite{Beze11} for special
cases of Dirichlet and Neumann boundary conditions. The VEVs of the field
squared and energy-momentum tensor in the region between two parallel plates
are discussed in section \ref{sec:VEVphi2}. Various special cases are
considered and the behavior of the VEVs in asymptotic regions of the
parameters is investigated. The Casimir forces acting on the plates are
studied in section \ref{sec:Forces}. Unlike to the case of the Minkowski
bulk, these forces are inhomogeneous and depend on the distance from the
string. Depending on the latter and on the boundary conditions, the presence
of the cosmic string can either increase or decrease the Casimir pressure.
The main results of the paper are summarized in section \ref{Conc}. In
appendix \ref{sec:Zeta} we present the evaluation of a more general
two-point object, the off-diagonal zeta function and also the local zeta
function.

\section{Problem setup and the heat kernel}

\label{sec:Setup}

We consider a massive scalar quantum field propagating in a $D$-dimensional
generalized cosmic string spacetime. By using the generalized cylindrical
coordinates with the cosmic string on the subspace defined by $r=0$, being $%
r\geqslant 0$ the radial polar coordinate, the corresponding metric tensor
is defined by the line element below:
\begin{equation}
ds^{2}=g_{ik}dx^{i}dx^{k}=-dt^{2}+dr^{2}+r^{2}d\varphi
^{2}+dz^{2}+\sum_{l=4}^{D-1}(dx^{l})^{2}\ .  \label{cs1}
\end{equation}%
The coordinate system reads: $x^{i}=(t,r,\varphi ,z,x^{l})$, with $\varphi
\in \lbrack 0,\ 2\pi /q]$, and $t,\ z,\ x^{l}\in (-\infty ,\ \infty )$. The
parameter $q$, smaller than unity, codifies the presence of the string. In a
4-dimensional spacetime, this parameter is related to the linear mass
density of the string, $\mu $, by $q^{-1}=1-4G\mu $, with $G$ being the
Newton gravitational constant. In this analysis we shall admit the presence
of extra coordinates, $x^{l}$, defined in an Euclidean $(D-4)$-dimensional
subspace.

For a scalar field propagating in an arbitrary curved spacetime the field
equation reads
\begin{equation}
\left( \Box -m^{2}-\xi R\right) \phi (x)=0\ ,  \label{eq1}
\end{equation}%
with $\Box $ denoting the covariant d'Alembertian operator and $R$ is the
scalar curvature. In (\ref{eq1}) we have introduced an arbitrary curvature
coupling $\xi $. The minimal coupling corresponds to $\xi =0$ and for the
conformal one $\xi =\xi _{c}=(D-2)/[4(D-1)]$. We shall assume that the field
obeys Robin boundary conditions
\begin{equation}
(1+\beta _{j}n_{j}^{\mu }\partial _{\mu })\phi (x)=0,\quad z=a_{j},
\label{Rob}
\end{equation}%
on the hypersurfaces orthogonal to the string and located at $z=a_{1}=0$ and
$z=a_{2}=a$. In (\ref{Rob}), $\beta _{j}$, $j=1,2$, are constants and $%
n_{j}^{\mu }$ is the inward pointing normal to the boundary at $z=a_{j}$. In
the region between the plates, $a_{1}\leqslant z\leqslant a_{2}$, one has $%
n_{j}^{\mu }=(-1)^{j-1}\delta _{3}^{\mu }$.

The Green function associated with a massive scalar field in a curved
spacetime obeys the second order differential equation
\begin{equation}
\left( \Box -m^{2}-\xi R\right) G(x,x^{\prime })=-\delta ^{D}(x,x^{\prime
})=-\frac{\delta ^{D}(x-x^{\prime })}{\sqrt{-g}}\ ,  \label{Green1}
\end{equation}%
where $\delta ^{D}(x,x^{\prime })$ represents the bidensity Dirac
distribution. This function can be obtained within the framework of the
Schwinger-DeWitt formalism as follows:
\begin{equation}
G(x,x^{\prime })=\int_{0}^{\infty }ds\ K(x,x^{\prime };s)\ ,  \label{heat}
\end{equation}%
where the heat kernel, $K(x,x^{\prime };s)$, is expressed in terms of a
complete set of normalized eigenfunctions of the operator defined in (\ref%
{eq1}):
\begin{equation}
K(x,x^{\prime };s)=\sum_{\sigma }\Phi _{\sigma }(x)\Phi _{\sigma }^{\ast
}(x^{\prime })e^{-s\sigma ^{2}}\ ,  \label{heat1}
\end{equation}%
with $\sigma ^{2}$ being the corresponding positively defined eigenvalue.

Writing
\begin{equation}
\left( \Box -m^{2}-\xi R\right) \Phi _{\sigma }(x)=-\sigma ^{2}\Phi _{\sigma
}(x)\ ,  \label{EqPhi}
\end{equation}%
in the spacetime defined by the line element (\ref{cs1}), a complete set of
normalized solutions in the region $a_{1}\leqslant z\leqslant a_{2}$ is
given by
\begin{equation}
\Phi _{\sigma }(x)=C_{\sigma }e^{i(nq\varphi +\mathbf{k}\cdot \mathbf{x}%
-\omega t)}J_{q|n|}(\lambda r)W(z)\ ,  \label{sol}
\end{equation}%
being $J_{\nu }(x)$ the Bessel function, $\mathbf{x}=(x^{4},\ldots ,x^{D-1})$%
, and%
\begin{equation}
W(z)=\cos \left[ k_{z}\left( z-a_{j}\right) +\gamma _{j}(k_{z})\right] .
\label{Wz}
\end{equation}%
For the quantum numbers in (\ref{sol}) one has $n=0,\pm 1,\pm 2,\ ...\ $, $%
\mathbf{k}$ $=(k_{4},\ldots ,k_{D-1})$, $-\infty <\omega ,\ k_{i}<+\infty $,
and $\lambda \geqslant 0$. From the boundary condition (\ref{Rob}) on the
plate at $z=a_{j}$, for the function $\gamma _{j}(k_{z})$ one obtains
\begin{equation}
e^{2i\gamma _{j}(k_{z})}=\frac{(-1)^{j}ik_{z}\beta _{j}+1}{%
(-1)^{j}ik_{z}\beta _{j}-1}.  \label{alfj}
\end{equation}%
From the boundary condition on the second plate it follows that the
eigenvalues for $k_{z}$ are solutions of the equation
\begin{equation}
\left( 1-b_{1}b_{2}y^{2}\right) \sin y-(b_{2}+b_{1})y\cos y=0.
\label{EigEq2}
\end{equation}%
where
\begin{equation}
y=k_{z}a,\;b_{j}=\beta _{j}/a.  \label{bj}
\end{equation}%
This equation coincides with the corresponding eigenvalue equation for two
parallel plates in Minkowski bulk \cite{Rome02} (for the scalar Casimir
densities for parallel plates with Robin boundary conditions on anti-de
Sitter, de Sitter and Friedmann-Robertson-Walker backgrounds see \cite%
{Saha05}). The equation (\ref{EigEq2}) has an infinite number of positive
roots which will be denoted by $y=y_{p}$, $p=1,2,\ldots $, and for the
corresponding eigenvalues of $k_{z}$ one has $k_{z}=y_{p}/a$. As a result,
the complete set of quantum numbers is specified by $(\omega ,\lambda ,n,p,%
\mathbf{k})$ and the corresponding positively defined eigenvalue is given by
\begin{equation}
\sigma ^{2}=\omega ^{2}+\lambda ^{2}+y_{p}^{2}/a^{2}+\mathbf{k}^{2}+m^{2}\ .
\label{sig2}
\end{equation}%
Note that, in addition to the real roots, depending on the values of $b_{1}$
and $b_{2}$, the equation (\ref{EigEq2}) may have one or two purely
imaginary roots (see \cite{Rome02}). Here, for simplicity of the further
discussion we will assume the values of the coefficients for which all the
roots are real.

The coefficient $C_{\sigma }$ in (\ref{sol}) is found by the normalization
condition
\begin{equation}
\sum_{\sigma }\Phi _{\sigma }(x)\Phi _{\sigma }^{\ast }(x^{\prime })=\delta
^{D}(x,x^{\prime })\ .  \label{Norm}
\end{equation}%
This gives
\begin{equation}
|C_{\sigma }|^{2}=\frac{2(2\pi )^{2-D}q\lambda /a}{1+\cos [y+2\tilde{\gamma}%
_{j}(y)]\sin (y)/y},  \label{Ck}
\end{equation}%
with $y=y_{p}$ and the function $\tilde{\gamma}_{j}(y)$ is defined by the
relation%
\begin{equation}
e^{2i\tilde{\gamma}_{j}(y)}=\frac{iyb_{j}-1}{iyb_{j}+1}\ ,\;\ j=1,\ 2\ .
\label{alfjtild}
\end{equation}

The next step is the evaluation of the heat kernel by using (\ref{heat1}).
On the base of (\ref{sol}) we have
\begin{eqnarray}
K(x,x^{\prime };s) &=&\frac{2q}{(2\pi )^{D-2}a}\int_{-\infty }^{\infty
}d\omega \int d\mathbf{k}\int_{0}^{\infty }d\lambda \lambda \sum_{n=-\infty
}^{\infty }\sum_{p=1}^{\infty }e^{i(qn\Delta \varphi +\mathbf{k}\cdot \Delta
\mathbf{x}-\omega \Delta t)}  \notag \\
&\times &\frac{J_{q|n|}(\lambda r)J_{q|n|}(\lambda r^{\prime
})W(z)W(z^{\prime })}{1+\cos [y_{p}+2\tilde{\gamma}_{j}(y_{p})]\sin
(y_{p})/y_{p}}e^{-s(\omega ^{2}+\lambda ^{2}+y_{p}^{2}/a^{2}+\mathbf{k}%
^{2}+m^{2})}\ ,  \label{hk1}
\end{eqnarray}%
where $\Delta \varphi =\varphi -\varphi ^{\prime }$, $\Delta t=t-t^{\prime }$%
, $\Delta \mathbf{x=x-x}^{\prime }$. After performing the integrals by using
the results from \cite{Grad,Prud86}, we obtain
\begin{equation}
K(x,x^{\prime };s)=\frac{2q\ e^{-\frac{\Delta \rho ^{2}}{4s}-sm^{2}}}{(4\pi
s)^{(D-1)/2}a}\sum_{n=-\infty }^{+\infty }e^{iqn\Delta \varphi
}I_{q|n|}\left( \frac{rr^{\prime }}{2s}\right) \sum_{p=1}^{\infty }\frac{%
W(z)W(z^{\prime })e^{-sy_{p}^{2}/a^{2}}}{1+\cos [y_{p}+2\tilde{\gamma}%
_{j}(y_{p})]\sin (y_{p})/y_{p}}\ ,  \label{heat2}
\end{equation}%
where $I_{\nu }(x)$ is the modified Bessel function \cite{Abra}, and
\begin{equation}
\Delta \rho ^{2}=r^{2}+r^{\prime }{}^{2}+\left( \Delta \mathbf{x}\right)
^{2}\ -\left( \Delta t\right) ^{2}.
\end{equation}%
Note that we can write%
\begin{equation}
W(z)W(z^{\prime })=\frac{1}{2}g_{j}(z,z^{\prime },k_{z}),  \label{Zpp}
\end{equation}%
where%
\begin{equation}
g_{j}(z,z^{\prime },k_{z})=\cos \left( k_{z}\Delta z\right) +\frac{1}{2}%
\sum_{\epsilon =\pm 1}e^{\epsilon ik_{z}|z+z^{\prime }-2a_{j}|}\frac{%
ik_{z}\beta _{j}-\epsilon }{ik_{z}\beta _{j}+\epsilon }.  \label{gj}
\end{equation}%
From here it follows that $g_{j}(z,z^{\prime },-k_{z})=g_{j}(z,z^{\prime
},k_{z})$.

The summation over the quantum number $n$ has been developed in \cite%
{Mello12}. The result is reproduced below:
\begin{equation}
\sum_{n=-\infty }^{\infty }e^{iqn\Delta \varphi }I_{q|n|}(v)=\frac{1}{q}%
\sum_{k}e^{v\cos (2k\pi /q-\Delta \varphi )}-\frac{1}{2\pi }\sum_{l=\pm
1}\int_{0}^{\infty }dy\frac{\sin (q\pi +lq\Delta \varphi )e^{-v\cosh y}}{%
\cosh (qy)-\cos (q\pi +lq\Delta \varphi )},  \label{sumform}
\end{equation}%
where the summation in the first term on the right hind side goes under the
condition
\begin{equation}
-q/2+q\Delta \varphi /(2\pi )\leqslant k\leqslant q/2+q\Delta \varphi /(2\pi
).  \label{SumCond}
\end{equation}%
If $\pm q/2+q\Delta \varphi /(2\pi )$ is an integer, then the corresponding
term in the first sum on the right-hand side of (\ref{sumform}) should be
taken with the coefficient 1/2. For integer values of $q$, formula (\ref%
{sumform}) reduces to the well-known result \cite{Prud86,Spin08}
\begin{equation}
\sum_{n=-\infty }^{\infty }e^{iqn\Delta \varphi }I_{qn}(v)=\frac{1}{q}%
\sum_{k=0}^{q-1}e^{v\cos (2k\pi /q-\Delta \varphi )}\ .  \label{SumFormSp}
\end{equation}%
By taking into account (\ref{sumform}), the heat kernel (\ref{heat2}) is
presented as%
\begin{eqnarray}
K(x,x^{\prime };s) &=&\frac{e^{-\frac{\Delta \rho ^{2}}{4s}-sm^{2}}}{(4\pi
s)^{(D-1)/2}a}\sum_{p=1}^{\infty }\frac{g_{j}(z,z^{\prime
},y_{p}/a)e^{-sy_{p}^{2}/a^{2}}}{1+\cos [y_{p}+2\tilde{\gamma}%
_{j}(y_{p})]\sin (y_{p})/y_{p}}  \notag \\
&\times & \left[ \sum_{k}e^{v\cos (2k\pi /q-\Delta \varphi )}-\frac{q}{2\pi }%
\sum_{l=\pm 1}\int_{0}^{\infty }dy\frac{\sin (q\pi +lq\Delta \varphi
)e^{-v\cosh y}}{\cosh (qy)-\cos (q\pi +lq\Delta \varphi )}\right] ,
\label{heat3}
\end{eqnarray}%
with $v=rr^{\prime }/(2s)$.

\section{Green function}

\label{sec:Green}

The Green function is evaluated by using (\ref{heat}) and (\ref{heat3}). The
integral over the variable $s$ is expressed in terms of the Macdonald
function $K_{\nu }(z)$ and the expression for the Green function takes the
form%
\begin{equation}
G(x,x^{\prime })=2(2\pi )^{\frac{1-D}{2}}\left[ \sum_{k}S(w_{k},x,x^{\prime
})-\frac{q}{2\pi }\int_{0}^{\infty }dy\,\sum_{l=\pm 1}\frac{\sin (q\pi
+lq\Delta \varphi )S(w_{y},x,x^{\prime })}{\cosh (qy)-\cos (q\pi +lq\Delta
\varphi )}\right] ,  \label{G1}
\end{equation}%
with the notation
\begin{equation}
S(w,x,x^{\prime })=\frac{1}{2a}\sum_{p=1}^{\infty }\frac{(m^{2}+k_{z}^{2})^{%
\frac{D-3}{2}}f_{\frac{D-3}{2}}(\sigma (w)\sqrt{m^{2}+k_{z}^{2}})}{1+\cos
[y_{p}+2\tilde{\gamma}_{j}(y_{p})]\sin (y_{p})/y_{p}}g_{j}(z,z^{\prime
},k_{z})\ ,  \label{Szz}
\end{equation}%
with $k_{z}=y_{p}/a$. Here we have introduced the function
\begin{equation}
f_{\nu }(x)=\frac{K_{\nu }(x)}{x^{\nu }},  \label{fnu}
\end{equation}%
and the notation%
\begin{equation}
\sigma (w)=\sqrt{-\Delta t^{2}+r^{2}+r^{\prime }{}^{2}+\Delta \mathbf{x}%
^{2}+2rr^{\prime }{w}}.  \label{sig}
\end{equation}%
Additionally, in (\ref{G1}) we have defined
\begin{equation}
w_{k}=-\cos (2k\pi /q-\Delta \varphi ),\;w_{y}=\cosh y.  \label{wk}
\end{equation}

In (\ref{Szz}), $y_{p}$ is given implicitly, as solutions of the
transcendental equation (\ref{EigEq2}), and that representation is not
convenient for the further evaluation of the VEVs in the coincidence limit.
An alternative representation is obtained by using a variant of the
generalized Abel-Plana formula \cite{Rome02,SahaRev}
\begin{eqnarray}
\sum_{p=1}^{\infty }\frac{\pi y_{p}f(y_{p})}{y_{p}+\cos [y_{p}+2\tilde{\gamma%
}_{j}(y_{p})]\sin y_{p}} &=&-\frac{\pi f(0)/2}{1-b_{2}-b_{1}}%
+\int_{0}^{\infty }duf(u)  \notag \\
&&+i\int_{0}^{\infty }du\frac{f(iu)-f(-iu)}{c_{1}(u)c_{2}(u)e^{2u}-1},
\label{sumfor}
\end{eqnarray}%
where, for the further convenience, the notation%
\begin{equation}
c_{j}(u)=\frac{b_{j}u-1}{b_{j}u+1}  \label{cj}
\end{equation}%
is introduced.

For the summation of the series in (\ref{Szz}) we take%
\begin{equation}
f(y)=(m^{2}+y^{2}/a^{2})^{\frac{D-3}{2}}f_{\frac{D-3}{2}}(\sigma (w)\sqrt{%
m^{2}+y^{2}/a^{2}})g_{j}(z,z^{\prime },y/a).  \label{fSum}
\end{equation}%
Note that one has $f(0)=0$. By taking into account that $g_{j}(z,z^{\prime
},-iu/a)=g_{j}(z,z^{\prime },iu/a)$, we see that $f(iu)-f(-iu)=0$ for $u<ma$
and
\begin{equation}
f(iu)-f(-iu)=-\pi ig_{j}(z,z^{\prime },iu/a)(u^{2}/a^{2}-m^{2})^{\frac{D-3}{2%
}}Z_{\frac{D-3}{2}}(\sigma (w)\sqrt{u^{2}/a^{2}-m^{2}}),  \label{fDiff}
\end{equation}%
for $u>ma$. Here
\begin{equation}
Z_{\nu }(x)=\frac{J_{\nu }(x)}{x^{\nu }},  \label{Znu}
\end{equation}%
and%
\begin{equation}
g_{j}(z,z^{\prime },iu/a)=\cosh \left( u\Delta z/a\right) +\frac{1}{2}%
\sum_{\epsilon =\pm 1}e^{-\epsilon u|z+z^{\prime }-2a_{j}|/a}\frac{%
ub_{j}+\epsilon }{ub_{j}-\epsilon }.  \label{gji}
\end{equation}%
Note that for the function $Z_{\nu }(x)$ one has the relation%
\begin{equation}
x^{2}Z_{\nu +1}(x)=2\nu Z_{\nu }(x)-Z_{\nu -1}(x).  \label{Zrec}
\end{equation}

Applying (\ref{sumfor}) with the function (\ref{fSum}) to the series in (\ref%
{Szz}) and by taking into account (\ref{gj}), the function $S(w,x,x^{\prime
})$ is decomposed as%
\begin{equation}
S(w,x,x^{\prime })=S_{0}(w,x,x^{\prime })+S_{j}(w,x,x^{\prime
})+S_{jj^{\prime }}(w,x,x^{\prime })\ ,  \label{Sdec}
\end{equation}%
where $j^{\prime }=1$ for $j=2$ and $j^{\prime }=2$ for $j=1$. The separate
terms are given by the expressions%
\begin{eqnarray}
S_{0}(w,x,x^{\prime }) &=&\frac{1}{2\pi }\int_{0}^{\infty
}du\,(m^{2}+u^{2})^{\frac{D-3}{2}}f_{\frac{D-3}{2}}(\sigma (w)\sqrt{%
m^{2}+u^{2}})\cos \left( u\Delta z\right) ,  \notag \\
S_{j}(w,x,x^{\prime }) &=&\frac{1}{4\pi }\int_{0}^{\infty
}du\,(m^{2}+u^{2})^{\frac{D-3}{2}}f_{\frac{D-3}{2}}(\sigma (w)\sqrt{%
m^{2}+u^{2}})  \notag \\
&&\times \sum_{\epsilon =\pm 1}e^{\epsilon iu|z+z^{\prime }-2a_{j}|}\frac{%
iu\beta _{j}-\epsilon }{iu\beta _{j}+\epsilon },  \notag \\
S_{jj^{\prime }}(w,x,x^{\prime }) &=&\frac{1}{2}\int_{m}^{\infty }du\frac{%
(u^{2}-m^{2})^{\frac{D-3}{2}}g_{j}(z,z^{\prime },iu)}{%
c_{1}(au)c_{2}(au)e^{2au}-1}Z_{\frac{D-3}{2}}(\sigma (w)\sqrt{u^{2}-m^{2}}).
\label{S0}
\end{eqnarray}%
The first two terms in the right-hand side of (\ref{Sdec}) come from the
first integral in (\ref{sumfor}). The integral in the expression $%
S_{0}(w,x,x^{\prime })$ is further evaluated with the result
\begin{equation}
S_{0}(w,x,x^{\prime })=\frac{m^{D-2}}{2\sqrt{2\pi }}f_{D/2-1}(m\sqrt{\sigma
^{2}(w)+(\Delta z)^{2}}).  \label{S0b}
\end{equation}%
In the part $S_{j}(w,x,x^{\prime })$ we rotate the integration contour in
the complex plane $u$ by the angle $\pi /2$ for the term $\epsilon =1$ and
by the angle $-\pi /2$ for $\epsilon =-1$. This gives%
\begin{equation}
S_{j}(w,x,x^{\prime })=\frac{1}{4}\int_{m}^{\infty }du\,(u^{2}-m^{2})^{\frac{%
D-3}{2}}Z_{\frac{D-3}{2}}(\sigma (w)\sqrt{u^{2}-m^{2}})e^{-u|z+z^{\prime
}-2a_{j}|}\frac{u\beta _{j}+1}{u\beta _{j}-1}.  \label{Sj}
\end{equation}

With the decomposition (\ref{Sdec}), the Green function (\ref{G1}) is
presented as%
\begin{equation}
G(x,x^{\prime })=G_{0}(x,x^{\prime })+G_{j}(x,x^{\prime })+G_{jj^{\prime
}}(x,x^{\prime }),  \label{Gdec2}
\end{equation}%
where%
\begin{equation}
G_{\alpha }(x,x^{\prime })=2(2\pi )^{\frac{1-D}{2}}\left[ \sum_{k}S_{\alpha
}(w_{k},x,x^{\prime })-\frac{q}{2\pi }\sum_{l=\pm 1}\int_{0}^{\infty }dy%
\frac{\sin (q\pi +lq\Delta \varphi )S_{\alpha }(w_{y},x,x^{\prime })}{\cosh
(qy)-\cos (q\pi +lq\Delta \varphi )}\right] ,  \label{Galf}
\end{equation}%
with $\alpha =0,j,jj^{\prime }$. Here, $G_{0}(x,x^{\prime })$ is the Green
function in the geometry without boundaries, the term $G_{j}(x,x^{\prime })$
is induced by the boundary at $z=a_{j}$ when the second boundary is absent
and the term $G_{jj^{\prime }}(x,x^{\prime })$ is induced if we add the
second boundary at $z=a_{j^{\prime }}$. The boundary-induced contribution,%
\begin{equation}
S_{\mathrm{b}}(w,x,x^{\prime })=S_{j}(w,x,x^{\prime })+S_{jj^{\prime
}}(w,x,x^{\prime }),  \label{Sb0}
\end{equation}%
can be combined in a single expression%
\begin{eqnarray}
S_{\mathrm{b}}(w,x,x^{\prime }) &=&\frac{1}{4}\int_{m}^{\infty }du\frac{%
(u^{2}-m^{2})^{(D-3)/2}}{c_{1}(au)c_{2}(au)e^{2au}-1}Z_{\frac{D-3}{2}%
}(\sigma (w)\sqrt{u^{2}-m^{2}})  \notag \\
&&\times \left[ 2\cosh \left( u\Delta z\right) +\sum_{j=1,2}e^{u|z+z^{\prime
}-2a_{j}|}c_{j}(au)\right] .  \label{Sb}
\end{eqnarray}%
Now, for the Green function we get the decomposition%
\begin{equation}
G(x,x^{\prime })=G_{0}(x,x^{\prime })+G_{\mathrm{b}}(x,x^{\prime }),
\label{Gdecb}
\end{equation}%
with the boundary-induced contribution%
\begin{equation}
G_{\mathrm{b}}(x,x^{\prime })=2(2\pi )^{\frac{1-D}{2}}\left[ \sum_{k}S_{%
\mathrm{b}}(w_{k},x,x^{\prime })-\frac{q}{2\pi }\sum_{j=\pm
1}\int_{0}^{\infty }dy\frac{\sin (q\pi +jq\Delta \varphi )S_{\mathrm{b}%
}(w_{y},x,x^{\prime })}{\cosh (qy)-\cos (q\pi +jq\Delta \varphi )}\right] .
\label{Gbn}
\end{equation}%
Note that%
\begin{equation*}
\sigma ^{2}(w_{0})+(\Delta z)^{2}=r^{2}+r^{\prime }{}^{2}+\Delta \mathbf{x}%
^{2}+2rr^{\prime }{\cos \Delta \varphi }+(\Delta z)^{2}-\Delta t^{2}.
\end{equation*}%
and the $k=0$ term in the expression (\ref{Galf}) for $G_{0}(x,x^{\prime })$
is the Green function in the boundary-free Minkowski spacetime. Hence, we
have obtained a representation for the Green function in which the
Minkowskian part is explicitly exhibited. This is important from the point
of view of the renormalization in the VEVs of the field squared and the
energy-momentum tensor. For points away from the cosmic string and
boundaries, the local geometry is the same as in the Minkowski spacetime
and, hence, the divergencies are the same as well. The renormalization in
the VEVs in the coincidence limit is reduced to the subtraction of the
Minkowskian part.

In the regions $z<a_{1}$ and $z>a_{2}$ the Green function is presented as
\begin{equation}
G(x,x^{\prime })=G_{0}(x,x^{\prime })+G_{j}(x,x^{\prime }),  \label{G1pl}
\end{equation}%
where $j=1$ ($j=2$) for the region $z<a_{1}$ ($z>a_{2}$). For special cases
of Dirichlet and Neumann boundary conditions, the integral in the expression
for (\ref{Sj}) for $S_{j}(w,x,x^{\prime })$ is expressed in terms of the
Macdonald function \cite{Grad} and we get%
\begin{equation}
S_{j}(w,x,x^{\prime })=\mp \frac{m^{D-2}}{2\sqrt{2\pi }}f_{\frac{D}{2}-1}(m%
\sqrt{\sigma ^{2}(w)+(z+z^{\prime }-2a_{j})^{2}}),  \label{SjDN}
\end{equation}%
where and in what follows the upper and lower signs correspond to Dirichlet
and Neumann conditions, respectively. In the case of Dirichlet boundary
condition, the expression (\ref{G1pl}), with (\ref{Galf}) and (\ref{SjDN})
coincides with the result of Ref. \cite{Beze11}.

For Dirichlet and Neumann boundary conditions and in the region between the
plates, an alternative representation of the function $S_{\mathrm{b}%
}(w,x,x^{\prime })$ is obtained from (\ref{Sb}) by using the expansion
\begin{equation}
\frac{1}{e^{2au}-1}=\sum_{n=1}^{\infty }e^{-2nau}.  \label{DNexp}
\end{equation}%
The integrals are evaluated by using the formula \cite{Prud86}%
\begin{equation}
\int_{0}^{\infty }dy\,y^{2\nu +1}Z_{\nu }(cy)\frac{e^{-b\sqrt{y^{2}+m^{2}}}}{%
\sqrt{y^{2}+m^{2}}}=\sqrt{\frac{2}{\pi }}m^{2\nu +1}f_{\nu +1/2}\left( m%
\sqrt{b^{2}+c^{2}}\right) .  \label{IntDN}
\end{equation}%
This leads to the result%
\begin{eqnarray}
S_{\mathrm{b}}(w,x,x^{\prime }) &=&\frac{m^{D-2}}{2\sqrt{2\pi }}%
\sum_{n=1}^{\infty }\sum_{j=1,2}\left[ f_{\frac{D}{2}-1}\left( m\sqrt{\sigma
^{2}(w)+\left( 2na-(-1)^{j}\Delta z\right) ^{2}}\right) \right.  \notag \\
&&\left. \mp f_{\frac{D}{2}-1}\left( m\sqrt{\sigma ^{2}(w)+\left(
2na-|z+z^{\prime }-2a_{j}|\right) ^{2}}\right) \right] .  \label{SbDN}
\end{eqnarray}%
A similar representation can be obtained for the function $S_{jj^{\prime
}}(w,x,x^{\prime })$. Combining (\ref{SbDN}) with (\ref{S0b}), the function $%
S(w,x,x^{\prime })$ in the expression for the Green function is presented in
the form%
\begin{eqnarray}
S(w,x,x^{\prime }) &=&\frac{m^{D-2}}{2\sqrt{2\pi }}\sum_{n=-\infty }^{\infty
}\left[ f_{\frac{D}{2}-1}\left( m\sqrt{\sigma ^{2}(w)+\left( 2na-\Delta
z\right) ^{2}}\right) \right.  \notag \\
&&\left. \mp f_{\frac{D}{2}-1}\left( m\sqrt{\sigma ^{2}(w)+\left(
2na+z+z^{\prime }-2a_{1}\right) ^{2}}\right) \right] .  \label{Sw}
\end{eqnarray}%
With this formula, the Green function\ $G(x,x^{\prime })$ in the region
between the plates is presented as an image sum of the Green functions in
the boundary-free geometry.

In Appendix \ref{sec:Zeta} we evaluate a more general two-point function,
namely, the off-diagonal zeta function. The latter is reduced to the Green
function for special value of the argument $s=1$. The local zeta function is
obtained from the off-diagonal zeta function in the coincidence limit of the
arguments corresponding to separated spacetime points.

\section{VEVs in the presence of a single plate}

\label{sec:VEV1pl}

This and the following sections will be devoted to the vacuum polarizations
effects induced by the boundaries. Two main calculations will be performed.
The evaluation of the VEV of the field squared, in the first place, followed
by the evaluation of the VEV of the energy-momentum tensor. Here we will
consider the VEVs in the presence of a single plate at $z=a_{j}$. The
corresponding Green function is presented as (\ref{G1pl}). For points away
from the boundary, the divergences in the coincidence limit $x^{\prime
}\rightarrow x$ are contained in the $k=0$ term of the expression (\ref{Galf}%
) for $G_{0}(x,x^{\prime })$. The latter corresponds to the Green function
in the boundary-free Minkowski spacetime. The renormalization is reduced to
the subtraction of the Minkowskian part.

\subsection{Field squared}

Taking the coincidence limit in (\ref{G1pl}) and omitting the Minkowskian
contribution, the VEV of the field squared is splitted as%
\begin{equation}
\langle \phi ^{2}\rangle =\langle \phi ^{2}\rangle _{\mathrm{cs}}+\langle
\phi ^{2}\rangle _{j}  \label{phi2}
\end{equation}%
where
\begin{equation}
\langle \phi ^{2}\rangle _{\mathrm{cs}}=\frac{2m^{D-2}}{(2\pi )^{\frac{D}{2}}%
}\left[ \sideset{}{'}{\sum}_{k=1}^{[q/2]}f_{\frac{D}{2}-1}(2mrs_{k})-\frac{%
q\sin (q\pi )}{\pi }\int_{0}^{\infty }dy\,\frac{f_{\frac{D}{2}-1}(2mr\cosh y)%
}{\cosh (2qy)-\cos (q\pi )}\right] ,  \label{phi2cs}
\end{equation}%
with $s_{k}=\sin (k\pi /q)$ is the renormalized VEV in the boundary-free
geometry. The prime on the summation sign in (\ref{phi2cs}) means that for
even values of $q$ the term with $k=[q/2]$ should be taken with the
coefficient 1/2. The part
\begin{equation}
\langle \phi ^{2}\rangle _{j}=(2\pi )^{\frac{1-D}{2}}\left[ %
\sideset{}{'}{\sum}_{k=0}^{[q/2]}U_{j}(s_{k},r,z)-\frac{q\sin (q\pi )}{\pi }%
\int_{0}^{\infty }dy\frac{U_{j}(\cosh y,r,z)}{\cosh (2qy)-\cos (q\pi )}%
\right] ,  \label{phi2jn}
\end{equation}%
is the boundary-induced contribution. In (\ref{phi2jn}), the prime on the
sign of the summation means that the terms $k=0$ and $k=[q/2]$ (for even
values of $q$) should be taken with the coefficient 1/2 and we have defined
the function
\begin{equation}
U_{j}(s,r,z)=\int_{m}^{\infty }du\,(u^{2}-m^{2})^{\frac{D-3}{2}}Z_{\frac{D-3%
}{2}}(2rs\sqrt{u^{2}-m^{2}})e^{-2u|z-a_{j}|}\frac{u\beta _{j}+1}{u\beta
_{j}-1}.  \label{Uj}
\end{equation}%
The $k=0$ term in (\ref{phi2jn}) coincides with the corresponding VEV\ for a
plate in the Minkowski bulk \cite{Rome02,SahaRev}:%
\begin{equation}
\langle \phi ^{2}\rangle _{j}^{(M)}=\frac{(4\pi )^{\frac{1-D}{2}}}{\Gamma
\left( \frac{D-1}{2}\right) }\int_{m}^{\infty }du\,(u^{2}-m^{2})^{\frac{D-3}{%
2}}e^{-2u|z-a_{j}|}\frac{u\beta _{j}+1}{u\beta _{j}-1}.  \label{phi2jM}
\end{equation}

For special cases of Dirichlet and Neuamnn boundary conditions, by using the
integral (\ref{IntDN}) we get%
\begin{equation}
U_{j}(s,r,z)=\mp \sqrt{\frac{2}{\pi }}m^{D-2}f_{\frac{D}{2}-1}(\chi ),
\label{UjND}
\end{equation}%
where
\begin{equation}
\chi =2m\sqrt{(z-a_{j})^{2}+r^{2}s^{2}},  \label{xi}
\end{equation}%
and the upper/lower sign corresponds to Dirichlet/Neaumann boundary
condition. The VEV\ (\ref{phi2jn}) with (\ref{UjND}) coincides with that
considered in \cite{Beze11}. The same expression is obtained by taking the
coincidence limit of $G_{j}(x,x^{\prime })$ with (\ref{Galf}) and (\ref{SjDN}%
).

Let us consider the asymptotic behavior of the VEV of the field squared in
limiting regions of the parameters. Near the cosmic string, $r\ll
m^{-1},|z-a_{j}|$, from (\ref{phi2cs}) for the boundary-free contribution to
the leading order we get%
\begin{equation}
\langle \phi ^{2}\rangle _{\mathrm{cs}}\approx \frac{\Gamma (D/2-1)g_{D-2}(q)%
}{2^{D-1}\pi ^{\frac{D}{2}}r^{D-2}},  \label{phi2cssm}
\end{equation}%
with the notation%
\begin{equation}
g_{n}(q)=\sideset{}{'}{\sum}_{k=1}^{[q/2]}s_{k}^{-n}-\frac{q\sin (q\pi )}{%
\pi }\int_{0}^{\infty }dy\frac{\cosh ^{-n}y}{\cosh (2qy)-\cos (q\pi )}.
\label{cnq}
\end{equation}%
For a massless field the result (\ref{phi2cssm}) is exact. For the
boundary-induced part in the limit $r\rightarrow 0$ we find%
\begin{equation}
\langle \phi ^{2}\rangle _{j}|_{r=0}=\langle \phi ^{2}\rangle _{j}^{(M)}
\left[ 1+2g_{0}(q)\right] .  \label{phi2jr0}
\end{equation}%
Taking in (\ref{sumform}) $\Delta \varphi =0$ and $v=0$, we can see that
\begin{equation}
g_{0}(q)=(q-1)/2,  \label{c0}
\end{equation}%
and, hence,%
\begin{equation}
\langle \phi ^{2}\rangle _{j}|_{r=0}=q\langle \phi ^{2}\rangle _{j}^{(M)}.
\label{phi2jr0b}
\end{equation}%
At large distances from the string, $r\gg |z-a_{j}|$, the topological part
in the boundary-induced contribution, $\langle \phi ^{2}\rangle _{j}-\langle
\phi ^{2}\rangle _{j}^{(M)}$, is suppressed by the factor $e^{-2mr\sin (\pi
/q)}$.

The boundary-induced VEV\ (\ref{phi2jn}) diverges on the boundary. This
divergence comes from the Minkowskian part and to the leading order
\begin{equation*}
\langle \phi ^{2}\rangle _{j}\approx \langle \phi ^{2}\rangle
_{j}^{(M)}\approx \frac{\left( 1-2\delta _{0\beta _{j}}\right) \Gamma \left(
D/2-1\right) }{(4\pi )^{\frac{D}{2}}|z-a_{j}|^{D-2}},
\end{equation*}%
for $|z-a_{j}|\ll r,m^{-1}$. For $r\neq 0$, the topological part $\langle
\phi ^{2}\rangle _{t}=\langle \phi ^{2}\rangle -\langle \phi ^{2}\rangle
_{j}^{(M)}$ is finite on the boundary, $z=a_{j}$. For Dirichlet and Neuamnn
boundary conditions this is obvious from (\ref{UjND}).

\subsection{Energy-momentum tensor}

Similar to the field squared, the VEV\ of the energy-momentum tensor is
presented as
\begin{equation}
\langle T_{\mu \nu }\rangle =\langle T_{\mu \nu }\rangle _{\mathrm{cs}%
}+\langle T_{\mu \nu }\rangle _{j},  \label{Tmu1pl}
\end{equation}%
where $\langle T_{\mu \nu }\rangle _{\mathrm{cs}}$ corresponds to the
geometry of the cosmic string without boundaries and $\langle T_{\mu \nu
}\rangle _{j}$ is induced by the boundary. Having the Green function and the
VEV of the field squared, the VEV\ of the energy-momentum tensor is
evaluated by using the formula%
\begin{equation}
\langle T_{\mu \nu }\rangle =\lim_{x^{\prime }\rightarrow x}\partial _{\mu
^{\prime }}\partial _{\nu }G(x,x^{\prime })+\left[ \left( \xi -{1}/{4}%
\right) g_{\mu \nu }\Box -\xi \nabla _{\mu }\nabla _{\nu }-\xi R_{\mu \nu }%
\right] \langle \phi ^{2}\rangle \ ,  \label{Tmuj}
\end{equation}%
where for the spacetime under consideration the Ricci tensor, $R_{\mu \nu }$%
, vanishes.

First let us consider the boundary-free part. By taking into account (\ref%
{Galf}) with $\alpha =0$ and (\ref{phi2cs}), we can see that the VEV $%
\langle T_{\mu \nu }\rangle _{\mathrm{cs}}$ is diagonal with the components
(no summation over $\mu $)%
\begin{equation}
\langle T_{\mu }^{\mu }\rangle _{\mathrm{cs}}=\frac{2m^{D}}{(2\pi )^{\frac{D%
}{2}}}\left[ \sum_{k=1}^{[q/2]}F^{(\mu )}(s_{k},2mrs_{k})-\frac{q\sin (q\pi )%
}{\pi }\int_{0}^{\infty }dy\frac{F^{(\mu )}(\cosh y,2mr\cosh y)}{\cosh
(2qy)-\cos (q\pi )}\right] ,  \label{Tmucs}
\end{equation}%
where%
\begin{eqnarray}
F^{(l)}(s,y) &=&\left( 4\xi -{1}\right) s^{2}y^{2}f_{\frac{D}{2}+1}(y)+\left[
1-2\left( 4\xi -{1}\right) s^{2}\right] f_{\frac{D}{2}}(y),  \notag \\
F^{(1)}(s,y) &=&\left( 1-4\xi s^{2}\right) f_{\frac{D}{2}}(y),  \notag \\
F^{(2)}(s,y) &=&\left( 1-4\xi s^{2}\right) \left[ f_{\frac{D}{2}}(y)-y^{2}f_{%
\frac{D}{2}+1}(y)\right] ,  \label{F2}
\end{eqnarray}%
with $l=0,3,\ldots ,D-1$. For integer values of $q$, (\ref{Tmucs}) is
reduced to the result given in \cite{Beze06sc}. In the case of a massless
field, by taking into account that $f_{\nu }(x)\approx 2^{\nu -1}\Gamma (\nu
)x^{-2\nu }$ for small $x$, one gets (no summation over $\mu $)%
\begin{equation}
\langle T_{\mu }^{\mu }\rangle _{\mathrm{cs}}=\frac{\Gamma (D/2)}{(4\pi )^{%
\frac{D}{2}}r^{D}}\left[ a_{\mu }^{(1)}g_{D-2}(q)+a_{\mu }^{(2)}g_{D}(q)%
\right] ,  \label{Tmusm0}
\end{equation}%
with the coefficients%
\begin{eqnarray}
a_{l}^{(1)} &=&\left( D-2\right) \left( 4\xi -{1}\right) ,\;a_{l}^{(2)}=1,
\notag \\
a_{1}^{(1)} &=&-4\xi ,\;a_{1}^{(2)}=1,  \notag \\
a_{2}^{(1)} &=&4\left( D-1\right) \xi ,\;a_{2}^{(2)}=1-D,  \label{a2}
\end{eqnarray}%
where $l=0,3,\ldots ,D-1$.

Now we turn to the boundary-induced contribution in the geometry of a single
plate at $z=a_{j}$. By taking into account the expression (\ref{Galf}) with $%
\alpha =j$ for the function $G_{j}(x,x^{\prime })$ and (\ref{phi2jn}), it is
presented in the form%
\begin{equation}
\langle T_{\mu \nu }\rangle _{j}=(2\pi )^{\frac{1-D}{2}}\left[ %
\sideset{}{'}{\sum}_{k=0}^{[q/2]}U_{\mu \nu }^{(j)}(s_{k},r,z)-\frac{q\sin
(q\pi )}{\pi }\int_{0}^{\infty }dy\frac{U_{\mu \nu }^{(j)}(\cosh y,r,z)}{%
\cosh (2qy)-\cos (q\pi )}\right] ,  \label{Tmuj1}
\end{equation}%
where the functions $U_{\mu \nu }^{(j)}(s,r,z)$ are defined by the relation
\begin{equation}
U_{\mu \nu }^{(j)}(s,r,z)=4\lim_{x^{\prime }\rightarrow x}\partial _{\mu
^{\prime }}\partial _{\nu }S_{j}(w,x,x^{\prime })+\left[ \left( \xi -{1}/{4}%
\right) g_{\mu \nu }\Box -\xi \nabla _{\mu }\nabla _{\nu }\right]
U_{j}(s,r,z)\ ,  \label{Umuja}
\end{equation}%
with $w=2s^{2}-1$. By using (\ref{S0}) for $S_{j}(w,x,x^{\prime })$ and (\ref%
{Uj}) we find the representation%
\begin{equation}
U_{\mu \nu }^{(j)}(s,r,z)=\int_{m}^{\infty }du\,(u^{2}-m^{2})^{\frac{D-1}{2}}%
\frac{u\beta _{j}+1}{u\beta _{j}-1}e^{-2u|z-a_{j}|}V_{\mu \nu }(u,s,r),
\label{Umuj}
\end{equation}%
where%
\begin{eqnarray}
e^{-2u|z-a_{j}|}V_{\mu \nu }(u,s,r) &=&\lim_{x^{\prime }\rightarrow
x}\partial _{\mu ^{\prime }}\partial _{\nu }Z_{\frac{D-3}{2}}(\sigma (w)%
\sqrt{u^{2}-m^{2}})\frac{e^{-u|z+z^{\prime }-2a_{j}|}}{u^{2}-m^{2}}  \notag
\\
&&+\left[ \left( \xi -{1}/{4}\right) g_{\mu \nu }\Box -\xi \nabla _{\mu
}\nabla _{\nu }\right] Z_{\frac{D-3}{2}}(\gamma )\frac{e^{-2u|z-a_{j}|}}{%
u^{2}-m^{2}}\ .  \label{Vmuj}
\end{eqnarray}%
and%
\begin{equation}
\gamma =2rs\sqrt{u^{2}-m^{2}}.  \label{gam}
\end{equation}%
The $k=0$ term in (\ref{Tmuj1}) gives the corresponding VEV induced by a
plate in Minkowski bulk.

After long but straightforward calculations, for the diagonal components $%
\langle T_{\mu }^{\mu }\rangle _{j}$ with $\mu \neq 2$ one finds (no
summation over $l$)%
\begin{eqnarray}
V_{l}^{l}(u,s,r) &=&\left[ 1+\left( 4\xi -{1}\right) (D-3)s^{2}\right] Z_{%
\frac{D-1}{2}}(\gamma )  \notag \\
&&+\left( 4\xi -{1}\right) \left( \frac{u^{2}}{u^{2}-m^{2}}-s^{2}\right) Z_{%
\frac{D-3}{2}}(\gamma ),  \notag \\
V_{1}^{1}(u,s,r) &=&\frac{\left( 4\xi -{1}\right) u^{2}}{u^{2}-m^{2}}Z_{%
\frac{D-3}{2}}(\gamma )+\left( 1-4\xi s^{2}\right) Z_{\frac{D-1}{2}}(\gamma
),  \notag \\
V_{3}^{3}(u,s,r) &=&\left( 4\xi -{1}\right) s^{2}\left[ (D-3)Z_{\frac{D-1}{2}%
}(\gamma )-Z_{\frac{D-3}{2}}(\gamma )\right] ,  \label{V33}
\end{eqnarray}%
where $l=0,4,\ldots ,D-1$. The only nonzero off-diagonal component is given
by the expression%
\begin{equation}
V_{3}^{1}(u,s,r)=2\left( 1-4\xi \right) \mathrm{sgn}\left( z-a_{j}\right)
rus^{2}Z_{(D-1)/2}(\gamma )\ ,  \label{V31}
\end{equation}%
where $\mathrm{sgn}\left( x\right) =\pm x/|x|$. As expected, the diagonal
components are symmetric with respect to the plate whereas the off-diagonal
component changes the sign. Note that for the off-diagonal component the
term $k=0$ in (\ref{Tmuj1}) vanishes. This corresponds to the fact that in
the Minkowski bulk the vacuum energy-momentum tensor is diagonal. All the
off-diagonal components of $\langle T_{\mu \nu }\rangle _{j}$, except the
components $\langle T_{13}\rangle _{j}=\langle T_{31}\rangle _{j}$, vanish.
This property is a direct consequence of the problem homogeneity with
respect to the coordinates $x^{i}$ with $i\neq 1,3$. Of course, that can
also be seen by a direct evaluation.

The remaining component $V_{2}^{2}(u,s,r)$ is most easily found by using the
covariant continuity equation $\nabla _{\nu }\langle T_{\mu }^{\nu }\rangle
_{j}=0$. For the geometry under consideration the latter is reduced to two
equations%
\begin{eqnarray}
\partial _{r}\left( r\langle T_{1}^{1}\rangle _{j}\right) +r\partial
_{z}\langle T_{1}^{3}\rangle _{j}-\langle T_{2}^{2}\rangle _{j} &=&0,  \notag
\\
\partial _{r}\left( r\langle T_{3}^{1}\rangle _{j}\right) +r\partial
_{z}\langle T_{3}^{3}\rangle _{j} &=&0,  \label{CovEq}
\end{eqnarray}%
where $\langle T_{1}^{3}\rangle _{j}=\langle T_{3}^{1}\rangle _{j}$. By
using (\ref{Tmuj1}) and (\ref{Umuj}), similar relations are found for the
functions $V_{\mu }^{\nu }(u,s,r)$:%
\begin{eqnarray}
\partial _{r}[rV_{1}^{1}(u,s,r)]-2\mathrm{sgn}\left( z-a_{j}\right)
urV_{1}^{3}(u,s,r)-V_{2}^{2}(u,s,r) &=&0,  \notag \\
\partial _{r}[rV_{3}^{1}(u,s,r)]+2(-1)^{j}urV_{3}^{3}(u,s,r) &=&0.
\label{CovV}
\end{eqnarray}%
First of all, we can check that the second of the relations is indeed obeyed
by the functions (\ref{V33}) and (\ref{V31}). For the evaluation of the
component $V_{2}^{2}(u,s,r)$ we use the first of the equations (\ref{CovV}):%
\begin{equation}
V_{2}^{2}(u,s,r)=\frac{\left( 4\xi -{1}\right) u^{2}}{u^{2}-m^{2}}Z_{\frac{%
D-3}{2}}(\gamma )+\left( 4\xi s^{2}-1\right) \left[ (D-2)Z_{\frac{D-1}{2}%
}(\gamma )-Z_{\frac{D-3}{2}}(\gamma )\right] .  \label{V22}
\end{equation}

For the trace we find the relation%
\begin{equation}
e^{-2u|z-a_{j}|}V_{\mu }^{\mu }(u,s,r)=\left[ (D-1)\left( \xi -\xi
_{c}\right) \Box -m^{2}\right] e^{-2u|z-a_{j}|}Z_{\frac{D-3}{2}}(\gamma ),
\label{TrV}
\end{equation}%
where $\xi _{c}=(D-2)/[4(D-1)]$ is the curvature coupling parameter for a
conformally coupled field. From (\ref{TrV}) it follows that for the
energy-momentum tensor one has the standard trace relation%
\begin{equation}
\langle T_{\mu }^{\mu }\rangle _{j}=\left[ (D-1)\left( \xi -\xi _{c}\right)
\Box -m^{2}\right] \langle \phi ^{2}\rangle _{j}.  \label{TrRel}
\end{equation}%
This is an additional check for the evaluation procedure presented above.
For Dirichlet and Neumann boundary conditions, by using (\ref{IntDN}) and%
\begin{equation}
\int_{m}^{\infty }du\,(u^{2}-m^{2})^{\nu +1}Z_{\nu }(\gamma
)e^{-2u|z-a_{j}|}=\sqrt{\frac{2}{\pi }}m^{D}\left[ \left( D-1\right) f_{\nu
+3/2}(\chi )-4m^{2}r^{2}s^{2}f_{\nu +5/2}(\chi )\right] ,  \label{Int2}
\end{equation}%
with $\gamma $ and $\chi $ defined by (\ref{gam}) and (\ref{xi}), we can see
that from (\ref{Tmuj1}) the expressions derived in \cite{Beze11} are
obtained. The result (\ref{Int2}) is obtained from (\ref{IntDN}) by using
the relations%
\begin{equation}
f_{\nu }^{\prime }(x)=-xf_{\nu +1}(x),\;f_{\nu -1}(x)=x^{2}f_{\nu
+1}(x)-2\nu f_{\nu }(x).  \label{Relf}
\end{equation}%
In particular, for the off-diagonal component we get%
\begin{equation}
U_{3}^{(j)1}(s,r,z)=\mp 4\sqrt{\frac{2}{\pi }}\left( 1-4\xi \right)
m^{D+2}\left( z-a_{j}\right) rs^{2}f_{\frac{D}{2}+1}(\chi ).  \label{U31DN}
\end{equation}%
where $\chi $ is defined by (\ref{xi}). In particular, we see that the
off-diagonal component\ $\langle T_{3}^{1}\rangle =\langle T_{3}^{1}\rangle
_{j}$ vanishes on the plate in the cases of Dirichlet and Neumann boundary
conditions.

Having in mind the application to the evaluation of the Casimir force (see
section \ref{sec:Forces} below), here we provide an alternative
representation for the function $U_{3}^{(j)1}(s,r,z)$. From (\ref{Umuj}), by
taking into account (\ref{V31}), one gets%
\begin{eqnarray}
U_{3}^{(j)1}(s,r,z) &=&U_{3}^{(\mathrm{N}j)1}(s,r,z)+4\left( 1-4\xi \right)
\mathrm{sgn}\left( z-a_{j}\right) rs^{2}  \notag \\
&&\times \int_{m}^{\infty }du\,u(u^{2}-m^{2})^{\frac{D-1}{2}}\frac{%
e^{-2u|z-a_{j}|}}{u\beta _{j}-1}Z_{\frac{D-1}{2}}(\gamma ),  \label{U31alt}
\end{eqnarray}%
where $U_{3}^{(\mathrm{N}j)1}(s,r,z)$ is the corresponding function for
Neumann boundary condition and is given by (\ref{U31DN}) with the lower
sign. For the transformation of the remaining part we use the integral
representation%
\begin{equation}
\frac{1}{1-u\beta _{j}}=\int_{0}^{\infty }dx\,e^{-\left( 1-u\beta
_{j}\right) x}.  \label{IntRep}
\end{equation}%
Substituting into (\ref{U31alt}) and changing the order of integrations, the
integral over $u$ is evaluated in terms of the Macdonald function and we find%
\begin{eqnarray}
U_{3}^{(j)1}(s,r,z) &=&U_{3}^{(\mathrm{N}j)1}(s,r,z)+8\sqrt{\frac{2}{\pi }}%
m^{D+2}\left( 1-4\xi \right) rs^{2}  \notag \\
&&\times \int_{0}^{\infty }dx\,e^{-x}\left[ z-a_{j}-\mathrm{sgn}\left(
z-a_{j}\right) \beta _{j}x/2\right]  \notag \\
&&\times f_{\frac{D}{2}+1}(2m\sqrt{(\left\vert z-a_{j}\right\vert -\beta
_{j}x/2)^{2}+r^{2}s^{2}}).  \label{U31alt2}
\end{eqnarray}%
In the special cases of Dirichlet and Neumann boundary conditions this
result is reduced to (\ref{U31DN}). The representation (\ref{U31alt2})
explicitly shows that the $\langle T_{3}^{1}\rangle _{j}$ is finite on the
plate. Note that for the representation (\ref{U31alt}) we are not allowed to
put directly $z=a_{j}$ in the integrand.

For points near the cosmic string, $r\ll m^{-1},|z-a_{j}|$, the dominant
contribution to the total VEV (\ref{Tmu1pl}) comes from the boundary-free
part and the leading terms in the diagonal components coincide with the VEV $%
\langle T_{\mu \nu }\rangle _{\mathrm{cs}}$ for a massless field given by (%
\ref{Tmusm0}). These terms diverge as $r^{-D}$. For $z\neq a_{j}$ the
boundary-induced contribution is finite on the string. Taking the limit $%
r\rightarrow 0$ in the expressions above we see that the VEVs are expressed
in terms of $g_{0}(q)$ and $g_{-2}(q)$. The function $g_{0}(q)$ has been
evaluated above and the function $g_{-2}(q)$ can be obtained from (\ref%
{sumform}) with $\Delta \varphi =0$, taking the derivative with respect to $%
v $ and then the limit $v\rightarrow 0$. In this way, we can see that
\begin{equation}
g_{-2}(q)=q/4.  \label{cm2}
\end{equation}%
For the diagonal components of the boundary-induced energy-momentum tensor
on the string one finds (no summation over $\mu $)%
\begin{equation}
\langle T_{\mu }^{\mu }\rangle _{j}|_{r=0}=q\langle T_{\mu }^{\mu }\rangle
_{j}^{(M)}-\frac{(4\pi )^{\frac{1-D}{2}}qd_{\mu }}{2\Gamma \left( \frac{D+1}{%
2}\right) }\int_{m}^{\infty }du\,(u^{2}-m^{2})^{\frac{D-1}{2}}\frac{u\beta
_{j}+1}{u\beta _{j}-1}e^{-2u|z-a_{j}|},  \label{Tmur0}
\end{equation}%
with the coefficients%
\begin{eqnarray}
d_{\mu } &=&4\xi -{1,\;\mu =0,3,\ldots ,D-1,}  \notag \\
d_{\mu } &=&2\xi {,\;\mu =1,2.}  \label{dmu}
\end{eqnarray}%
The VEV\ for a plate in Minkowski spacetime is given by \cite{Rome02,SahaRev}
\begin{eqnarray}
\langle T_{\mu }^{\nu }\rangle _{j}^{(M)} &=&\frac{(4\pi )^{\frac{1-D}{2}%
}\delta _{\mu }^{\nu }}{2\Gamma \left( \frac{D+1}{2}\right) }%
\int_{m}^{\infty }du\,(u^{2}-m^{2})^{\frac{D-3}{2}}\frac{u\beta _{j}+1}{%
u\beta _{j}-1}  \notag \\
&&\times e^{-2u|z-a_{j}|}\left[ 4(D-1)\left( \xi -\xi _{c}\right) u^{2}-m^{2}%
\right] ,  \label{TmuM}
\end{eqnarray}%
for $\mu \neq 3$ and $\langle T_{3}^{3}\rangle _{j}^{(M)}=0$. The leading
term in the asymptotic expansion for the off-diagonal component near the
string is given by
\begin{equation}
\langle T_{3}^{1}\rangle _{j}\approx \mathrm{sgn}\left( z-a_{j}\right) \frac{%
(4\pi )^{\frac{1-D}{2}}qr}{2\Gamma \left( \frac{D+1}{2}\right) }\left(
1-4\xi \right) \int_{m}^{\infty }du\,u(u^{2}-m^{2})^{\frac{D-1}{2}}\frac{%
u\beta _{j}+1}{u\beta _{j}-1}e^{-2u|z-a_{j}|},  \label{T31str}
\end{equation}%
and this component vanishes on the string.

The boundary-induced VEV $\langle T_{\mu \nu }\rangle _{j}$ diverges on the
boundary. For points outside the string, $r\neq 0$, the divergences are the
same as those for Minkowski bulk and the topological part induced by the
string, $\langle T_{\mu \nu }\rangle -\langle T_{\mu \nu }\rangle _{j}^{(M)}$
is finite on the boundary. Consequently, to the leading order for $\mu \neq
3 $ and $|z-a_{j}|\ll m^{-1},r$ one has (no summation over $\mu $)%
\begin{equation}
\langle T_{\mu }^{\mu }\rangle \approx \langle T_{\mu }^{\mu }\rangle
_{j}^{(M)}\approx \frac{2(D-1)\Gamma \left( \frac{D}{2}\right) \delta _{\mu
}^{\nu }}{(4\pi )^{\frac{D}{2}}|z-a_{j}|^{D}}\left( \xi -\xi _{c}\right)
\left( 1-2\delta _{0\beta _{j}}\right) ,  \label{TmuNear}
\end{equation}%
for $\mu \neq 3$. At large distances from the string, $r\gg |z-a_{j}|,m^{-1}$%
, the topological part in the boundary-induced contribution $\langle T_{\mu
\nu }\rangle _{j}-\langle T_{\mu \nu }\rangle _{j}^{(M)}$ is suppressed by
the factor $e^{-2mr\sin (\pi /q)}$.

\section{VEVs in the region between the plates}

\label{sec:VEVphi2}

Now we turn to the case of two plates and will consider the region between
them, $a_{1}\leqslant z\leqslant a_{2}$. The VEVs in the regions $z\leqslant
a_{1}$ and $z\geqslant a_{2}$ are given by the expression from the previous
sections with $j=1$ and $j=2$ respectively.

\subsection{Field squared}

The VEV of the field squared is formally given by evaluating the Green
function at the coincidence limit. In this analysis the complete Green
function is given by the sum (\ref{Gdecb}). Omitting the part corresponding
to the boundary-free Minkowski spacetime, we obtain
\begin{equation}
\langle \phi ^{2}\rangle =\langle \phi ^{2}\rangle _{\mathrm{cs}}+\langle
\phi ^{2}\rangle _{\mathrm{b}}\ ,  \label{phi2sum}
\end{equation}%
where the boundary-free contribution is given by (\ref{phi2cs}). For the
boundary-induced part one gets the expression%
\begin{equation}
\langle \phi ^{2}\rangle _{\mathrm{b}}=(2\pi )^{\frac{1-D}{2}}\left[ %
\sideset{}{'}{\sum}_{k=0}^{[q/2]}U(s_{k},r,z)-\frac{q\sin (q\pi )}{\pi }%
\int_{0}^{\infty }dy\frac{U(\cosh y,r,z)}{\cosh (2qy)-\cos (q\pi )}\right] ,
\label{phi2b}
\end{equation}%
with the function
\begin{equation}
U(s,r,z)=\int_{m}^{\infty }du\frac{(u^{2}-m^{2})^{\frac{D-3}{2}}g(u,z)}{%
c_{1}(au)c_{2}(au)e^{2au}-1}Z_{\frac{D-3}{2}}(2rs\sqrt{u^{2}-m^{2}}).
\label{U}
\end{equation}%
In (\ref{U}) we have defined%
\begin{equation}
g(u,z)=2+\sum_{j=1,2}e^{2u|z-a_{j}|}c_{j}(au).  \label{guz}
\end{equation}%
The $k=0$ term in (\ref{phi2b}) is the corresponding VEV in the region
between two plates on the Minkowski bulk:%
\begin{equation}
\langle \phi ^{2}\rangle _{\mathrm{b}}^{(M)}=\frac{U(0,r,z)}{2(2\pi )^{\frac{%
D-1}{2}}}=\frac{(4\pi )^{-\frac{D-1}{2}}}{\Gamma \left( \frac{D-1}{2}\right)
}\int_{ma}^{\infty }du\frac{(u^{2}-m^{2})^{\frac{D-3}{2}}g(u,z)}{%
c_{1}(au)c_{2}(au)e^{2u}-1}.  \label{phi2bM}
\end{equation}

In the special cases of Dirichlet and Neumann boundary conditions,
equivalent expression for (\ref{U}) is obtained from (\ref{Gbn}) with the
function (\ref{SbDN}):%
\begin{equation}
U(s,r,z)=\sqrt{\frac{2}{\pi }}m^{D-2}\sum_{n=1}^{\infty }\left[ 2f_{\frac{D}{%
2}-1}\left( \chi _{n}\right) \mp \sum_{j=1,2}f_{\frac{D}{2}-1}\left( \chi
_{jn}\right) \right] ,  \label{UDN}
\end{equation}%
where the upper and lower signs correspond to Dirichlet and Neumann boundary
conditions and
\begin{eqnarray}
\chi _{n} &=&2m\sqrt{n^{2}a^{2}+r^{2}s^{2}},  \notag \\
\chi _{jn} &=&2m\sqrt{\left( na-|z-a_{j}|\right) ^{2}+r^{2}s^{2}}.
\label{xin}
\end{eqnarray}%
For a massless field, taking the limit $m\rightarrow 0$ in (\ref{UDN}), one
gets%
\begin{equation}
U(s,r,z)=\frac{\Gamma \left( \frac{D}{2}-1\right) }{2^{\frac{D-1}{2}}\sqrt{%
\pi }}\sum_{n=1}^{\infty }\left[ \frac{2}{(n^{2}a^{2}+r^{2}s^{2})^{\frac{D}{2%
}-1}}\mp \sum_{j=1,2}\frac{1}{[\left( na-|z-a_{j}|\right) ^{2}+r^{2}s^{2}]^{%
\frac{D}{2}-1}}\right] .  \label{UDNm0}
\end{equation}

For $z\neq a_{1},a_{2}$, the boundary-induced part $\langle \phi ^{2}\rangle
_{\mathrm{b}}$ is finite on the string:%
\begin{equation}
\langle \phi ^{2}\rangle _{\mathrm{b}}|_{r=0}=q\langle \phi ^{2}\rangle _{%
\mathrm{b}}^{(M)}.  \label{phi2br0}
\end{equation}%
Near the string, the boundary-free part behaves as (\ref{phi2cssm}) and it
dominates in the total VEV. At large distances from the cosmic string the
leading contribution to (\ref{phi2b}) comes from the term $k=0$ that
coincides with the corresponding VEV\ for Minkowski bulk, $\langle \phi
^{2}\rangle _{\mathrm{b}}^{(M)}$. For a massive field, under the conditions $%
r\gg a,m^{-1}$, the topological contribution in the boundary-induced part, $%
\langle \phi ^{2}\rangle _{\mathrm{b}}-\langle \phi ^{2}\rangle _{\mathrm{b}%
}^{(M)}$, is suppressed by the factor $e^{-2mr\sin (\pi /q)}$. Comparing
with (\ref{phi2cs}), we see that this contribution is of the same order as
the boundary-free topological part $\langle \phi ^{2}\rangle _{\mathrm{cs}}$%
. For a massless field and for $r\gg a$, the topological contribution $%
\langle \phi ^{2}\rangle _{\mathrm{b}}-\langle \phi ^{2}\rangle _{\mathrm{b}%
}^{(M)}$ in the case of non-Dirichlet boundary conditions behaves as $%
r^{3-D}/a$. By taking into account that $\langle \phi ^{2}\rangle _{\mathrm{%
cs}}\propto 1/r^{D-2}$, we conclude that for a massless field and at large
distances the topological part in the VEV\ of the field squared is dominated
by the boundary-induced contribution.

An alternative form for the VEV\ of the field squared is obtained by using
the representation (\ref{Gdec2}):%
\begin{equation}
\langle \phi ^{2}\rangle =\langle \phi ^{2}\rangle _{\mathrm{cs}}+\langle
\phi ^{2}\rangle _{j}+\langle \phi ^{2}\rangle _{jj^{\prime }},
\label{phi22pl}
\end{equation}%
where the second boundary-induced part is given as%
\begin{equation}
\langle \phi ^{2}\rangle _{jj^{\prime }}=(2\pi )^{\frac{1-D}{2}}\left[
\sum_{k=0}^{[q/2]}U_{jj^{\prime }}(s_{k},r,z)-\frac{q\sin (q\pi )}{\pi }%
\int_{0}^{\infty }dy\frac{U_{jj^{\prime }}(\cosh y,r,z)}{\cosh (2qy)-\cos
(q\pi )}\right] ,  \label{phi2jj}
\end{equation}%
with the function%
\begin{equation}
U_{jj^{\prime }}(s,r,z)=\int_{m}^{\infty }du\frac{(u^{2}-m^{2})^{\frac{D-3}{2%
}}g_{j}(u,z)}{c_{1}(au)c_{2}(au)e^{2au}-1}Z_{\frac{D-3}{2}}(2rs\sqrt{%
u^{2}-m^{2}}),  \label{Ujjn}
\end{equation}%
where%
\begin{equation}
g_{j}(u,z)=2+e^{2u|z-a_{j}|}c_{j}(au)+\frac{e^{-2u|z-a_{j}|}}{c_{j}(au)}.
\label{gj1pl}
\end{equation}%
Note that the contribution $\langle \phi ^{2}\rangle _{jj^{\prime }}$ is
finite on the plate at $z=a_{j}$. For Dirichlet boundary condition it
vanishes on that plate. In the special cases of Dirichlet and Neumann
boundary conditions, alternative expressions for $U_{jj^{\prime }}(s,r,z)$,
similar to (\ref{UDN}), are obtained by using the expansion (\ref{DNexp}).

\subsection{Energy-momentum tensor}

Following the same line of investigation, in this section we are interested
in the evaluation of the contribution induced by the boundaries in the VEV
of the energy-momentum tensor. Similar to the case of the field squared, in
the region between the plates the energy-momentum tensor is presented in the
splitted form,
\begin{equation}
\langle T_{\mu \nu }\rangle =\langle T_{\mu \nu }\rangle _{\mathrm{cs}%
}+\langle T_{\mu \nu }\rangle _{\mathrm{b}},  \label{EMTdec}
\end{equation}%
where the boundary-free part $\langle T_{\mu \nu }\rangle _{\mathrm{cs}}$ is
given by (\ref{Tmucs}). In order to evaluate the boundary-induced
contribution we will use the analog of the formula (\ref{Tmuj}) for that
contribution. By using the expression (\ref{Gbn}) for the boundary-induced
part in the Green functions, the VEV of the energy-momentum tensor is
presented as
\begin{equation}
\langle T_{\nu }^{\mu }\rangle _{\mathrm{b}}=(2\pi )^{\frac{1-D}{2}}\left[ %
\sideset{}{'}{\sum}_{k=0}^{[q/2]}U_{\nu }^{\mu }(s_{k},r,z)-\frac{q\sin
(q\pi )}{\pi }\int_{0}^{\infty }dy\frac{U_{\nu }^{\mu }(\cosh y,r,z)}{\cosh
(2qy)-\cos (q\pi )}\right] .  \label{Tmub}
\end{equation}%
For the diagonal components one has the function (no summation over $\mu $)%
\begin{equation}
U_{\mu }^{\mu }(s,r,z)=\int_{m}^{\infty }du\frac{(u^{2}-m^{2})^{\frac{D-1}{2}%
}}{c_{1}(au)c_{2}(au)e^{2au}-1}\left[ 2V_{0\mu }^{\mu }(u,s,r)+V_{\mu }^{\mu
}(u,s,r)\sum_{j=1,2}e^{2u|z-a_{j}|}c_{j}(au)\right] ,  \label{Umu}
\end{equation}%
where the functions $V_{\mu }^{\mu }(u,s,r)$ are defined by (\ref{V33}), (%
\ref{V22}) and (no summation over $l$)%
\begin{eqnarray}
V_{0l}^{l}(u,s,r) &=&\left[ s^{2}\left( 4\xi -1\right) \left( D-3\right) +1%
\right] Z_{\frac{D-1}{2}}(\gamma )+s^{2}\left( 1-4\xi \right) Z_{\frac{D-3}{2%
}}(\gamma ),  \notag \\
V_{01}^{1}(u,s,r) &=&\left( 1-4\xi s^{2}\right) Z_{\frac{D-1}{2}}(\gamma ),
\notag \\
V_{02}^{2}(u,s,r) &=&\left( 1-4\xi s^{2}\right) \left[ (2-D)Z_{\frac{D-1}{2}%
}(\gamma )+Z_{\frac{D-3}{2}}(\gamma )\right] ,  \notag \\
V_{03}^{3}(u,s,r) &=&\left( 4\xi -1\right) s^{2}\left[ (D-3)Z_{\frac{D-1}{2}%
}(\gamma )-Z_{\frac{D-3}{2}}(\gamma )\right] -u^{2}\frac{Z_{\frac{D-3}{2}%
}(\gamma )}{u^{2}-m^{2}},  \label{V033}
\end{eqnarray}%
where $l=0,4,\ldots ,D-1$ and $\gamma $ is defined by (\ref{gam}).

The azimuthal component of the energy-momentum tensor is most easily found
from the analog of the first equation in (\ref{CovEq}). For the only nonzero
off-diagonal component one gets%
\begin{equation}
U_{3}^{1}(s,r,z)=2\left( 1-4\xi \right) rs^{2}\int_{m}^{\infty }du\frac{%
u(u^{2}-m^{2})^{\frac{D-1}{2}}Z_{\frac{D-1}{2}}(\gamma )}{%
c_{1}(au)c_{2}(au)e^{2au}-1}\sum_{j=1,2}(-1)^{j}e^{2u|z-a_{j}|}c_{j}(au).
\label{U31}
\end{equation}%
Now we can check that the boundary-induced VEV obeys the second equation in (%
\ref{CovEq}) and the trace relation (\ref{TrRel}). The $k=0$ term in (\ref%
{Tmub}) gives the VEV for parallel plates in the Minkowski bulk: $\langle
T_{\nu }^{\mu }\rangle _{\mathrm{b}}^{(M)}=(2\pi )^{\frac{1-D}{2}}U_{\nu
}^{\mu }(0,0,z)/2$. The latter does not depend on the radial coordinate and
the off-diagonal component vanishes.

For Dirichlet and Neumann boundary conditions, alternative expressions for
the VEVs are obtained by using the expansion (\ref{DNexp}). The integral
over $u$ in (\ref{Umu}) is expressed in terms of the Macdonald function. As
a result, the function appearing in the expression (\ref{Tmub}) for the
diagonal components is presented in the form (no summation over $\mu $)%
\begin{equation}
U_{\mu }^{\mu }(s,r,z)=\sqrt{\frac{2}{\pi }}m^{D}\sum_{n=1}^{\infty }\left[
2W_{0\mu }(s,\chi _{n})\mp \sum_{j=\pm 1}W_{\mu }(s,\chi _{jn})\right] ,
\label{UmuDNm}
\end{equation}%
where the upper and lower signs correspond to Dirichlet and Neumann boundary
conditions, respectively, and $\chi _{n}$, $\chi _{jn}$ are given by (\ref%
{xin}). The functions in (\ref{UmuDNm}) are defined by the relations%
\begin{eqnarray}
W_{0l}(s,x) &=&\left[ 2\left( 1-4\xi \right) s^{2}+1\right] f_{\frac{D}{2}%
}\left( x\right) -s^{2}\left( 1-4\xi \right) \left( 2mrs\right) ^{2}f_{\frac{%
D}{2}+1}\left( x\right) ,  \notag \\
W_{01}(s,x) &=&\left( 1-4\xi s^{2}\right) f_{\frac{D}{2}}\left( x\right) ,
\notag \\
W_{02}(s,x) &=&\left( 1-4\xi s^{2}\right) \left[ f_{\frac{D}{2}}\left(
x\right) -\left( 2mrs\right) ^{2}f_{\frac{D}{2}+1}\left( x\right) \right] ,
\notag \\
W_{03}(s,x) &=&W_{00}^{0}(s,x)+\left[ \left( 2mrs\right) ^{2}-x^{2}\right]
f_{\frac{D}{2}+1}\left( x\right) ,  \label{W303}
\end{eqnarray}%
and%
\begin{eqnarray}
W_{l}(s,x) &=&\left[ \left( 1-4\xi \right) \left( 2s^{2}+1\right) +1\right]
f_{\frac{D}{2}}\left( x\right)  \notag \\
&&+\left( 4\xi -{1}\right) \left[ x^{2}+\left( s^{2}-1\right) \left(
2mrs\right) ^{2}\right] f_{\frac{D}{2}+1}\left( x\right) ,  \notag \\
W_{1}(s,x) &=&\left[ 2-4\xi \left( s^{2}+1\right) \right] f_{\frac{D}{2}%
}\left( x\right) +\left( 4\xi -{1}\right) \left[ x^{2}-\left( 2mrs\right)
^{2}\right] f_{\frac{D}{2}+1}\left( x\right) ,  \notag \\
W_{2}(s,x) &=&W_{1}^{1}(s,x)+\left( 4\xi s^{2}-{1}\right) \left( 2mrs\right)
^{2}f_{\frac{D}{2}+1}\left( x\right) ,  \notag \\
W_{3}(s,x) &=&\left( 1-4\xi \right) s^{2}\left[ 2f_{\frac{D}{2}}\left(
x\right) -\left( 2mrs\right) ^{2}f_{\frac{D}{2}+1}\left( x\right) \right] ,
\label{W33}
\end{eqnarray}%
with $l=0,4,\ldots ,D-1$. For the off-diagonal component one gets%
\begin{equation}
U_{3}^{1}(s,r,z)=\mp 4\sqrt{\frac{2}{\pi }}m^{D+2}\left( 1-4\xi \right)
rs^{2}\sum_{n=1}^{\infty }\sum_{j=1,2}(-1)^{j}\left( na-|z-a_{j}|\right) f_{%
\frac{D}{2}+1}(\chi _{jn}).  \label{U31DNb}
\end{equation}%
Note that the off-diagonal components vanishes on the plates for Dirichlet
and Neumann boundary conditions.

For a massless field and for Dirichlet and Neumann boundary conditions, the
expressions for the VEVs are obtained from (\ref{UmuDNm})-(\ref{U31DNb}) by
making use of the asymptotic expression $f_{\nu }(x)\approx 2^{\nu -1}\Gamma
\left( \nu \right) x^{-2\nu }$ for $x\ll 1$. The diagonal components are
presented as (no summation over $\mu $)%
\begin{equation}
U_{\mu }^{\mu }(s,r,z)=\frac{\Gamma \left( D/2\right) a^{-D}}{2^{(D+1)/2}%
\sqrt{\pi }}\sum_{n=1}^{\infty }\left[ \frac{2W_{0\mu }^{(0)}(s,na/(rs))}{%
[n^{2}+\left( sr/a\right) ^{2}]^{D/2}}\mp \sum_{j=\pm 1}\frac{W_{\mu
}^{(0)}(s,(na-|z-a_{j}|)/(rs))}{[\left( n-|z-a_{j}|/a\right) ^{2}+\left(
sr/a\right) ^{2}]^{D/2}}\right] ,  \label{UmuDNm0}
\end{equation}%
with the functions%
\begin{eqnarray}
W_{0l}^{(0)}(s,x) &=&1+\left( 1-4\xi \right) s^{2}\left( 2-\frac{D}{1+x^{2}}%
\right) ,  \notag \\
W_{01}^{(0)}(s,x) &=&1-4\xi s^{2},  \notag \\
W_{02}^{(0)}(s,x) &=&\left( 1-4\xi s^{2}\right) \left( 1-\frac{D}{1+x^{2}}%
\right) ,  \notag \\
W_{03}^{(0)}(s,x) &=&W_{00}^{(0)}(s,x)-\frac{Dx^{2}}{1+x^{2}},  \label{W003}
\end{eqnarray}%
with $l=0,4,\ldots ,D-1$, and%
\begin{eqnarray}
W_{l}^{(0)}(s,x) &=&1+\left( 4\xi -{1}\right) \left( D-1-2s^{2}+D\frac{%
s^{2}-1}{1+x^{2}}\right) ,  \notag \\
W_{1}^{(0)}(s,x) &=&2-4\xi \left( s^{2}+1\right) +Dx^{2}\frac{4\xi -{1}}{%
1+x^{2}},  \notag \\
W_{2}^{(0)}(s,x) &=&W_{1}^{(0)}(s,x)+D\frac{4\xi s^{2}-{1}}{1+x^{2}},  \notag
\\
W_{3}^{(0)}(s,x) &=&\left( 1-4\xi \right) s^{2}\left( 2-\frac{D}{1+x^{2}}%
\right) .  \label{W03}
\end{eqnarray}%
For the off-diagonal component one gets%
\begin{equation}
U_{3}^{1}(s,r,z)=\mp \frac{D\left( 1-4\xi \right) rs^{2}}{2^{\frac{D+1}{2}}%
\sqrt{\pi }a^{D+1}}\Gamma \left( \frac{D}{2}\right) \sum_{n=1}^{\infty
}\sum_{j=1,2}\frac{(-1)^{j}\left( n-|z-a_{j}|/a\right) }{[\left(
n-|z-a_{j}|/a\right) ^{2}+\left( sr/a\right) ^{2}]^{D/2+1}}.  \label{U31DNm0}
\end{equation}

An alternative representation for general Robin boundary conditions is
obtained by using the decomposition (\ref{Gdec2}) for the Green function:%
\begin{equation}
\langle T_{\mu \nu }\rangle =\langle T_{\mu \nu }\rangle _{\mathrm{cs}%
}+\langle T_{\mu \nu }\rangle _{j}+\langle T_{\mu \nu }\rangle _{jj^{\prime
}},  \label{Tmualt}
\end{equation}%
where the second boundary induced contribution is given by the expression%
\begin{equation}
\langle T_{\nu }^{\mu }\rangle _{jj^{\prime }}=(2\pi )^{\frac{1-D}{2}}\left[ %
\sideset{}{'}{\sum}_{k=0}^{[q/2]}U_{\nu }^{(jj^{\prime })\mu }(s_{k},r,z)-%
\frac{q\sin (q\pi )}{\pi }\int_{0}^{\infty }dy\frac{U_{\nu }^{(jj^{\prime
})\mu }(\cosh y,r,z)}{\cosh (2qy)-\cos (q\pi )}\right] .  \label{Tmujj}
\end{equation}%
The diagonal components of the function in (\ref{Tmujj}) are defined as%
\begin{eqnarray}
U_{\mu }^{(jj^{\prime })\mu }(s,r,z) &=&\int_{m}^{\infty }du\frac{%
(u^{2}-m^{2})^{\frac{D-1}{2}}}{c_{1}(au)c_{2}(au)e^{2au}-1}  \notag \\
&&\times \left\{ 2V_{0\mu }^{\mu }(u,s,r)+V_{\mu }^{\mu }(u,s,r)\left[
e^{2u|z-a_{j}|}c_{j}(au)+\frac{e^{-2u|z-a_{j}|}}{c_{j}(au)}\right] \right\} .
\label{Umujj}
\end{eqnarray}%
For the off-diagonal component one has%
\begin{eqnarray}
U_{3}^{(jj^{\prime })1}(s,r,z) &=&2(-1)^{j}\left( 1-4\xi \right)
rs^{2}\int_{m}^{\infty }du\frac{u(u^{2}-m^{2})^{(D-1)/2}}{%
c_{1}(au)c_{2}(au)e^{2au}-1}  \notag \\
&&\times Z_{(D-1)/2}(\gamma )\left[ e^{2u|z-a_{j}|}c_{j}(au)-\frac{%
e^{-2u|z-a_{j}|}}{c_{j}(au)}\right] ,  \label{U31jj}
\end{eqnarray}%
The second boundary induced contribution $\langle T_{\mu \nu }\rangle
_{jj^{\prime }}$ is finite on the plate at $z=a_{j}$.

For Dirichlet and Neumann boundary conditions, an equivalent expression for
the function $U_{\mu }^{(jj^{\prime })\mu }(s,r,z)$ is obtained in a way
similar to that we have used for (\ref{UmuDNm}) (no summation over $\mu $):%
\begin{equation}
U_{\mu }^{(jj^{\prime })\mu }=\sqrt{\frac{2}{\pi }}m^{D}\sum_{n=1}^{\infty }%
\left[ 2W_{0\mu }(s,\chi _{n})\mp \sum_{l=\pm 1}W_{\mu }(s,\chi _{jn}^{(l)})%
\right] ,  \label{UmujjDN}
\end{equation}%
with $\chi _{jn}^{(l)}=2m\sqrt{\left( na-l|z-a_{j}|\right) ^{2}+r^{2}s^{2}}$
and with the functions (\ref{W303}) and (\ref{W33}). For the off-diagonal
component one gets%
\begin{equation}
U_{3}^{(jj^{\prime })1}(s,r,z)=\mp 4\sqrt{\frac{2}{\pi }}m^{D+2}(-1)^{j}%
\left( 1-4\xi \right) rs^{2}\sum_{n=1}^{\infty }\sum_{l=\pm 1}l\left(
na-l|z-a_{j}|\right) f_{\frac{D}{2}+1}(\chi _{jn}^{(l)}).  \label{Ujj31DN}
\end{equation}%
The latter vanishes on the plate $z=a_{j}$.

For points near the string, $r\ll m^{-1},|z-a_{j}|$, the boundary-free part
dominates in (\ref{EMTdec}). For a massive field the leading term in the
corresponding asymptotic expansion is given by (\ref{Tmusm0}). For points $%
z\neq a_{j}$, the boundary-induced contribution is finite on the string and
is expressed in terms of $g_{0}(q)$ and $g_{-2}(q)$. By using (\ref{c0}) and
(\ref{cm2}) for the diagonal components we find (no summation over $\mu $):%
\begin{equation}
\langle T_{\mu }^{\mu }\rangle _{\mathrm{b}}|_{r=0}=q\langle T_{\mu }^{\mu
}\rangle _{\mathrm{b}}^{(M)}-\frac{(4\pi )^{\frac{1-D}{2}}qd_{\mu }}{2\Gamma
\left( \frac{D+1}{2}\right) }\int_{m}^{\infty }du\frac{(u^{2}-m^{2})^{\frac{%
D-1}{2}}}{c_{1}(au)c_{2}(au)e^{2au}-1}\left[ 1+%
\sum_{j=1,2}e^{2u|z-a_{j}|}c_{j}(au)\right] ,  \label{Tmur02pl}
\end{equation}%
where the coefficients $d_{\mu }$ are given by (\ref{dmu}). The off-diagonal
component linearly vanishes on the string:%
\begin{equation}
\langle T_{3}^{1}\rangle _{\mathrm{b}}\approx \frac{(4\pi )^{\frac{1-D}{2}%
}q\left( 1-4\xi \right) r}{2\Gamma \left( \frac{D+1}{2}\right) }%
\int_{m}^{\infty }du\frac{u(u^{2}-m^{2})^{\frac{D-1}{2}}}{%
c_{1}(au)c_{2}(au)e^{2au}-1}\sum_{j=1,2}(-1)^{j}e^{2u|z-a_{j}|}c_{j}(au),
\label{T13r0}
\end{equation}%
for $r\rightarrow 0$.

The leading term in the asymptotic expansion of $\langle T_{\mu \nu }\rangle
$ at large distances from the string is given by the term $k=0$ in (\ref%
{Tmub}) and it coincides with the corresponding VEV\ for Minkowski bulk, $%
\langle T_{\nu }^{\mu }\rangle _{\mathrm{b}}^{(M)}$. Assuming that $r\gg
a,m^{-1}$, the topological part for a massive field in the boundary-induced
contribution, $\langle T_{\nu }^{\mu }\rangle _{\mathrm{b}}-\langle T_{\nu
}^{\mu }\rangle _{\mathrm{b}}^{(M)}$, is suppressed by the factor $%
e^{-2mr\sin (\pi /q)}$. The same is the case for the boundary-free
topological term $\langle T_{\nu }^{\mu }\rangle _{\mathrm{cs}}$. For a
massless field in the region $r\gg a$, the contribution $\langle T_{\nu
}^{\mu }\rangle _{\mathrm{b}}-\langle T_{\nu }^{\mu }\rangle _{\mathrm{b}%
}^{(M)}$ decays like $r^{1-D}/a$. Under the same conditions for the
boundary-free part one has $\langle T_{\nu }^{\mu }\rangle _{\mathrm{cs}%
}\propto 1/r^{D}$, and the dominant contribution comes from the
boundary-induced topological part.

\section{The Casimir forces}

\label{sec:Forces}

The $\mu $th component of the force acting on the surface element $dS$ of
the plate at $z=a_{j}$ is given by $-\langle T_{\nu }^{\mu }\rangle
_{z=a_{j}+0}n_{(+)j}^{\nu }dS$ in the region $z\geqslant a_{j}+0$ and $%
-\langle T_{\nu }^{\mu }\rangle _{z=a_{j}-0}n_{(-)j}^{\nu }dS$ in the region
$z\leqslant a_{j}-0$, where $n_{(\pm )j}^{\nu }=\pm \delta _{3}^{\nu }$. For
the resulting force we get
\begin{equation}
dF_{(j)}^{\mu }=\langle T_{3}^{\mu }\rangle |_{z=a_{j}+0}^{z=a_{j}-0}dS.
\label{dFj}
\end{equation}%
Due to the nonzero off-diagonal component $\langle T_{3}^{1}\rangle $, in
addition to the normal component $dF_{(j)}^{3}$, this force has nonzero
component parallel to the boundary (shear force), $dF_{(j)}^{1}$. First we
will consider the normal force.

\subsection{Normal force}

For the normal force acting on the plate at $z=a_{j}$ one has $%
dF_{(j)}^{3}=\langle T_{3}^{3}\rangle |_{z=a_{j}+0}^{z=a_{j}-0}dS$. For $%
\langle T_{3}^{3}\rangle $ we have the decomposition (\ref{Tmualt}) in the
region between the plates and (\ref{Tmu1pl}) in the remaining regions. The
parts $\langle T_{3}^{3}\rangle _{\mathrm{cs}}$ and $\langle
T_{3}^{3}\rangle _{j}$ are the same on the left and right-hand sides of the
plate and they do not contribute to the net force. The nonzero contribution
comes from the term $\langle T_{3}^{3}\rangle _{jj^{\prime }}$ in the region
between the plates. Hence, for the vacuum effective pressure on the plate $%
z=a_{j}$ one gets%
\begin{equation}
P_{j}=\langle T_{3}^{3}\rangle _{jj^{\prime }}|_{z=a_{j}}=(2\pi )^{\frac{1-D%
}{2}}\left[ \sideset{}{'}{\sum}_{k=0}^{[q/2]}F_{j}(s_{k},r)-\frac{q\sin
(q\pi )}{\pi }\int_{0}^{\infty }dy\frac{F_{j}(\cosh y,r)}{\cosh (2qy)-\cos
(q\pi )}\right] ,  \label{Pj}
\end{equation}%
where%
\begin{eqnarray}
F_{j}(s,r) &=&-\int_{m}^{\infty }du\frac{(u^{2}-m^{2})^{\frac{D-1}{2}}}{%
c_{1}(au)c_{2}(au)e^{2au}-1}\left\{ 2u^{2}\frac{Z_{\frac{D-3}{2}}(\gamma )}{%
u^{2}-m^{2}}\right.  \notag \\
&&\left. +\left( 1-4\xi \right) s^{2}\left[ (D-3)Z_{\frac{D-1}{2}}(\gamma
)-Z_{\frac{D-3}{2}}(\gamma )\right] h_{j}(au)\right\} ,  \label{Fj}
\end{eqnarray}%
with $\gamma $ given by (\ref{gam}), and%
\begin{equation}
h_{j}(au)=2+c_{j}(au)+\frac{1}{c_{j}(au)}.  \label{gjF}
\end{equation}%
The pressures (\ref{Pj}) with $j=1$ and $j=2$ act on the sides $z=a_{1}+0$
and $z=a_{2}-0$ of the plates, respectively. The corresponding forces are
attractive for $P_{j}<0$ and repulsive for $P_{j}>0$. Note that $%
dF_{(j)}^{3}=(-1)^{j}P_{j}dS$. The $k=0$ term in (\ref{Pj}) corresponds to
the Casimir pressure for plates in the Minkowski bulk \cite{Rome02,SahaRev}:%
\begin{equation}
P_{j}^{(M)}=-\frac{2(4\pi )^{\frac{1-D}{2}}}{\Gamma \left( \frac{D-1}{2}%
\right) }\int_{m}^{\infty }du\frac{u^{2}(u^{2}-m^{2})^{\frac{D-3}{2}}}{%
c_{1}(au)c_{2}(au)e^{2au}-1}.  \label{PjM}
\end{equation}%
For special cases of Dirichlet and Neumann boundary conditions the Casimir
forces coincide on Minkowski bulk and they are attractive. In the presence
of the cosmic string, these forces, in general, are different.

Let us consider the behaviour of the Casimir forces in the asymptotic
regions of the parameters. For points on the string, $r=0$, by taking into
account that $Z_{\nu }(0)=2^{-\nu }/\Gamma (\nu +1)$ and using (\ref{c0}), (%
\ref{cm2}), one finds%
\begin{equation}
P_{j}|_{r=0}=qP_{j}^{(M)}-\frac{(4\pi )^{\frac{1-D}{2}}\left( 4\xi -1\right)
q}{2\Gamma \left( \frac{D+1}{2}\right) a^{D}}\int_{ma}^{\infty }dx\frac{%
(x^{2}-m^{2}a^{2})^{\frac{D-1}{2}}}{c_{1}(x)c_{2}(x)e^{2x}-1}h_{j}(x),
\label{Pjr0}
\end{equation}%
where $P_{j}^{(M)}$ is given by (\ref{PjM}). For Dirichlet boundary
condition the second term in the right hand side (\ref{Pjr0}) vanishes and $%
P_{j}|_{r=0}=qP_{j}^{(M)}$. At large distances from the string, $r\gg
a,m^{-1}$, the leading term in the asymptotic expansion of $P_{j}$ coincides
with the corresponding quantity for the plates in Minkowski bulk, given by (%
\ref{PjM}). The topological contribution is suppressed by the factor $%
e^{-2mr\sin (\pi /q)}$ for a massive field and decays as $1/r^{D-1}$ for a
massless field.

In the case of Dirichlet boundary condition, an alternative expression for
the Casimir pressure is obtained by using the expansion (\ref{DNexp}). With
this expansion, the function $F_{j}(s,r)$ is presented as%
\begin{equation}
F_{j}^{(D)}(s,r)=-\frac{1}{2}\sum_{n=1}^{\infty }\frac{\partial ^{2}}{%
\partial \left( na\right) ^{2}}\int_{0}^{\infty }dy\,y^{D-2}Z_{\frac{D-3}{2}%
}(2rsy)\frac{e^{-2na\sqrt{y^{2}+m^{2}}}}{\sqrt{y^{2}+m^{2}}}.  \label{FjD}
\end{equation}%
The integral is evaluated by using the formula (\ref{IntDN}). In addition,
by making use the relations (\ref{Relf}), we get the representation%
\begin{equation}
F_{j}^{(D)}(s,r)=2\sqrt{\frac{2}{\pi }}m^{D}\sum_{n=1}^{\infty }\left[ f_{%
\frac{D}{2}}\left( \chi _{n}\right) -\left( 2mna\right) ^{2}f_{\frac{D}{2}%
+1}\left( \chi _{n}\right) \right] .  \label{FjD1}
\end{equation}%
For a massless field this gives%
\begin{equation}
F_{j}^{(D)}(s,r)=\frac{\Gamma \left( D/2\right) }{2^{\frac{D-1}{2}}\sqrt{\pi
}}\sum_{n=1}^{\infty }\frac{r^{2}s^{2}-\left( D-1\right) n^{2}a^{2}}{\left(
n^{2}a^{2}+r^{2}s^{2}\right) ^{\frac{D}{2}+1}}.  \label{FjDm0}
\end{equation}%
Note that for Dirichlet boundary condition the Casimir pressure do not
depend on the curvature coupling parameter.

In a similar way, for Neumann boundary condition the function (\ref{Fj}) is
presented as%
\begin{equation}
F_{j}^{(N)}(s,r)=F_{j}^{(D)}(s,r)+4\left( 4\xi -1\right) s^{2}\sqrt{\frac{2}{%
\pi }}m^{D}\sum_{n=1}^{\infty }\left[ \left( 2mrs\right) ^{2}f_{\frac{D}{2}%
+1}(\chi _{n})-2f_{\frac{D}{2}}\left( \chi _{n}\right) \right] .  \label{FjN}
\end{equation}%
For the Casimir pressure at the location of the string, by using (\ref{c0})
and (\ref{cm2}), we get%
\begin{equation*}
P_{j}|_{r=0}=qP_{j}^{(M)}+qm^{D}\frac{4\left( 1-4\xi \right) }{(2\pi )^{%
\frac{D}{2}}}\sum_{n=1}^{\infty }f_{\frac{D}{2}}\left( 2mna\right) ,
\end{equation*}%
where%
\begin{equation}
P_{j}^{(M)}=\frac{2m^{D}}{(2\pi )^{\frac{D}{2}}}\sum_{n=1}^{\infty }\left[
f_{\frac{D}{2}}\left( 2mna\right) -\left( 2mna\right) ^{2}f_{\frac{D}{2}%
+1}\left( 2mna\right) \right] ,  \label{PjMD}
\end{equation}%
is the Casimir pressure in Minkowski spacetime for Dirichlet and Neuamann
boundary conditions. As seen, unlike the Dirichlet case, the Casimir
pressure for Neuamnn boundary condition depends on the curvature coupling
parameter $\xi $.

For a massless scalar field and for Neumann boundary condition the
expression (\ref{FjN}) is simplified to%
\begin{equation}
F_{j}^{(N)}(s,r)=F_{j}^{(D)}(s,r)+\frac{4\Gamma \left( D/2\right) \left(
4\xi -1\right) }{2^{\frac{D-1}{2}}\sqrt{\pi }}s^{2}\sum_{n=1}^{\infty }\frac{%
(D/2-1)r^{2}s^{2}-n^{2}a^{2}}{\left( n^{2}a^{2}+r^{2}s^{2}\right) ^{\frac{D}{%
2}+1}},  \label{FjNm0}
\end{equation}%
with $F_{j}^{(D)}(s,r)$ from (\ref{FjDm0}). For the Casimir pressure at the
location of the string this gives
\begin{equation}
P_{j}|_{r=0}=qP_{j}^{(M)}\left( 1+2\frac{4\xi -1}{D-1}\right) ,
\label{Pjr0N}
\end{equation}%
with the Minkowskian Casimir pressure%
\begin{equation}
P_{j}^{(M)}=-\frac{\left( D-1\right) \Gamma \left( D/2\right) }{(4\pi )^{%
\frac{D}{2}}a^{D}}\zeta (D),  \label{PjMm0}
\end{equation}%
where $\zeta (x)$ is the Riemann zeta function. For even values of $D$ the
series in (\ref{FjDm0}) and (\ref{FjNm0}) are expressed in terms of
elementary functions by using the relation%
\begin{equation}
\sum_{n=1}^{\infty }\frac{1}{\pi ^{2}n^{2}+b^{2}}=\frac{b\coth b-1}{2b^{2}}
\label{Ser1}
\end{equation}%
and its derivatives with respect to $b$. For example, in $D=4$ one gets%
\begin{eqnarray}
F_{j}^{(D)}(s,r) &=&\frac{b^{3}\cosh b\sinh ^{-3}b-1}{4\sqrt{2\pi }r^{4}s^{4}%
},  \notag \\
F_{j}^{(N)}(s,r) &=&F_{j}^{(D)}(s,r)+\frac{\xi -1/4}{\sqrt{2\pi }r^{4}s^{2}}%
\left[ \frac{b^{2}}{\sinh ^{2}b}+b\coth b\left( 1+\frac{2b^{2}}{\sinh ^{2}b}%
\right) -4\right] ,  \label{FjDN}
\end{eqnarray}%
where $b=\pi rs/a$. For plates in Minkowski bulk the Casimir forces for
Dirichlet and Neumann boundary conditions coincide and for a massless field
we have%
\begin{equation}
P_{j}^{(M)}=\frac{F_{j}^{(D)}(0,r)}{2(2\pi )^{(D-1)/2}}=-\left( D-1\right)
\frac{\Gamma \left( D/2\right) \zeta (D)}{(4\pi )^{D/2}a^{D}}.  \label{PjMDN}
\end{equation}

\begin{figure}[tbph]
\begin{center}
\begin{tabular}{cc}
\epsfig{figure=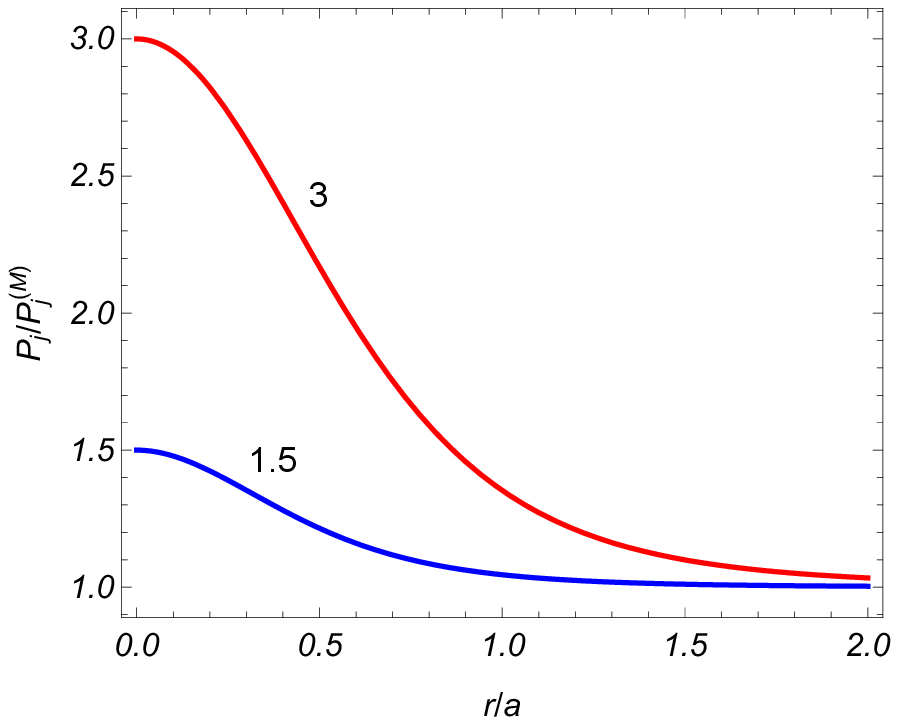,width=7.cm,height=5.5cm} & \quad %
\epsfig{figure=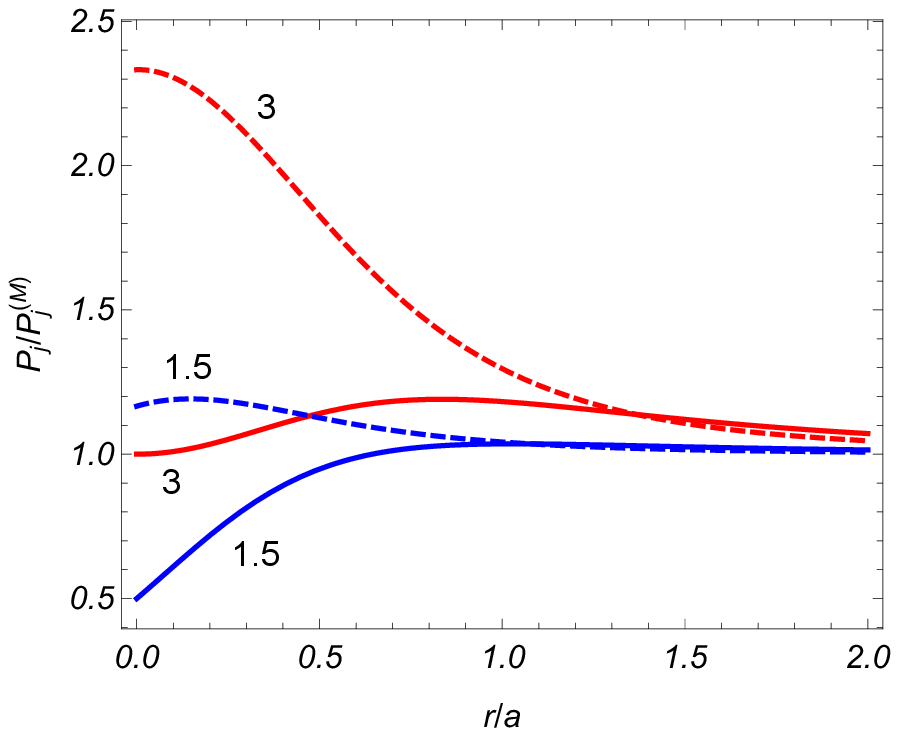,width=7.cm,height=5.5cm}%
\end{tabular}%
\end{center}
\caption{The ratio of the Casimir pressure for a massless scalar field to
the corresponding quantity in the Minkowski bulk versus the distance from
the string. The left and right panels are for Dirichlet and Neumann boundary
conditions respectively. The numbers near the curves are the values of the
parameter $q$. On the right panel, the full and dashed curves correspond to
minimally and conformally coupled scalars, respectively.}
\label{fig1}
\end{figure}

In figure \ref{fig1} we have plotted the ratio of the Casimir pressure to
the corresponding pressure in Minkowski bulk, $P_{j}^{(M)}$, as a function
of the distance from the axis of the cosmic string. The graphs are plotted
for a massless field. The left/right panel corresponds to Dirichlet/Neumann
boundary condition and the numbers near the curves correspond to the values
of the parameter $q$. In the right panel, the full/dashed lines correspond
to minimally/conformally coupled scalar fields. For Dirichlet boundary
condition the Casimir forces in the cases of minimal and conformal couplings
are the same. Note that for the considered example from (\ref{PjMDN}) one
has $P_{j}^{(M)}=-\pi ^{2}a^{-4}/480$. Recall that, at the location of the
string, $r=0$, one has $P_{j}/P_{j}^{(M)}=q$ for Dirichlet boundary
condition and the relation (\ref{Pjr0N}) for Neumann boundary condition.

Another special case corresponds to Dirichlet boundary condition on the
plate $z=a_{1}$ ($\beta _{1}=0$) and Neumann boundary condition on the
second plate $z=a_{2}$ ($\beta _{2}=\infty $). The corresponding expressions
for the function $F_{j}(s,r)$ is obtained from (\ref{Fj}) taking $%
c_{j}(au)=(-1)^{j}$, $h_{1}(au)=0$ and $h_{2}(au)=4$. We can obtain an
equivalent representations by using the expansion for the function $%
(e^{2au}+1)^{-1}$ which is the analog of (\ref{DNexp}). The expressions for $%
F_{1}(s,r)$ are obtained from the formulas (\ref{FjD1}) and (\ref{FjDm0})
for the function $F_{j}^{(D)}(s,r)$ adding the factor $(-1)^{n}$ in the
expression under the sign of the summation. The expressions for $F_{2}(s,r)$
are obtained from (\ref{FjN}) and (\ref{FjNm0}) in a similar way. For even
values of $D$ and for a massless field the series over $n$ is summed taking
the derivatives of the relation%
\begin{equation}
\sum_{n=1}^{\infty }\frac{(-1)^{n}}{\pi ^{2}n^{2}+b^{2}}=\frac{b/\sinh b-1}{%
2b^{2}}.  \label{Ser2}
\end{equation}%
In particular, for $D=4$ we can show that%
\begin{eqnarray}
F_{1}^{(DN)}(s,r) &=&\frac{1}{8\sqrt{2\pi }r^{4}s^{4}}\left( b^{3}\frac{%
2+\sinh ^{2}b}{\sinh ^{3}b}-2\right) ,  \notag \\
F_{2}^{(DN)}(s,r) &=&F_{1}^{(DN)}(s,r)+\frac{\xi -1/4}{\sqrt{2\pi }r^{4}s^{2}%
}\left[ \frac{b}{\sinh b}\left( \frac{2b^{2}}{\sinh ^{2}b}+b\coth
b+b^{2}+1\right) -4\right] ,  \label{F12DN}
\end{eqnarray}%
with the same $b$ as in (\ref{FjDN}).

\begin{figure}[tbph]
\begin{center}
\begin{tabular}{cc}
\epsfig{figure=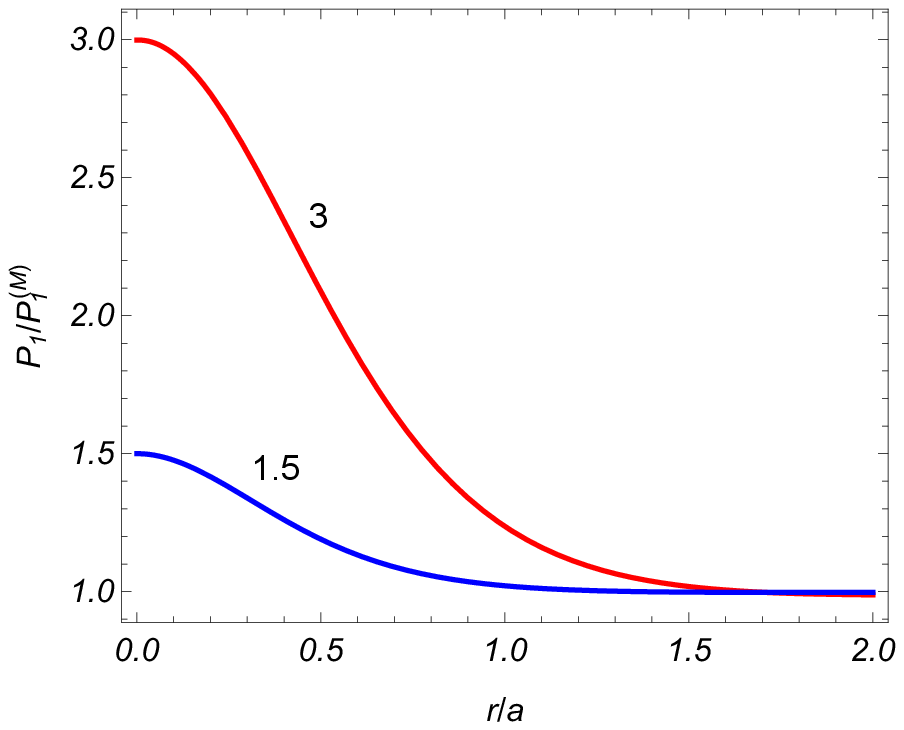,width=7.cm,height=5.5cm} & \quad %
\epsfig{figure=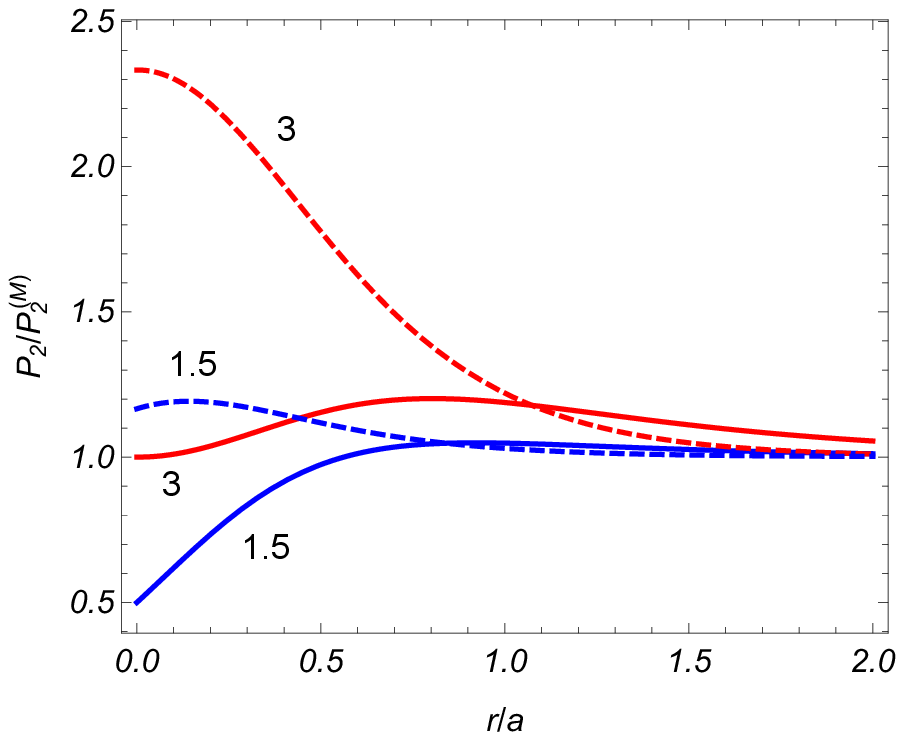,width=7.cm,height=5.5cm}%
\end{tabular}%
\end{center}
\caption{The ratio of the Casimir pressures for a massless scalar field in $%
D=4$ cosmic string and Minkowski backgrounds in the case of Dirichlet
boundary condition at $z=a_{1}$ and Neumann boundary condition at $z=a_{2}$.
The left (right) panel presents the pressure on the plate $z=a_{1}$ ($%
z=a_{2} $). On the right panel, the full and dashed curves correspond to
minimally and conformally coupled scalars, respectively. The numbers near
the curves are the values of the parameter $q$. }
\label{fig2}
\end{figure}

For the model with $D=4$, figure \ref{fig2} displays the ratio $%
P_{j}/P_{j}^{(M)}$ versus $r/a$ for Dirichlet boundary condition on the
plate $z=a_{1}$ and Neumann boundary condition on the plate $z=a_{2}$. The
graphs are plotted for a massless field and for two values of the parameter $%
q$ (numbers near the curves). The left and right panels correspond to the
plates $z=a_{1}$ and $z=a_{2}$, respectively. For the right panel, the
full/dashed lines correspond to minimally/conformally coupled scalar fields.
For the considered example we have $P_{j}^{(M)}=7\pi ^{2}a^{-4}/3840$ and
the corresponding forces are repulsive.

\subsection{Shear force}

As it has been emphasized above, in the problem at hand in addition to the
normal Casimir force one has a nonzero shear force along the radial
direction, $dF_{(j)}^{1}=f_{(j)}dS$, where $f_{(j)}$ is the shear force per
unit surface of the plate at $z=z_{j}$. The latter is given by $%
f_{(j)}=\langle T_{3}^{1}\rangle |_{z=a_{j}+0}^{z=a_{j}-0}$ (note that the
limiting transitions $r\rightarrow 0$ and $z\rightarrow a_{j}\pm 0$ are not
commutative).

Let us start with the case of a single plate at $z=z_{j}$. The corresponding
shear forces acting on the sides $z=a_{j}-0$ and $z=a_{j}+0$ coincide and by
using (\ref{Tmuj1}) with the function $U_{3}^{(j)1}(s,r,z)$ from (\ref%
{U31alt}) for $r\neq 0$ one gets%
\begin{equation}
f_{j}^{(1)}=\frac{4\beta _{j}\left( 1-4\xi \right) }{(2\pi )^{D/2}r}\left[ %
\sideset{}{'}{\sum}_{k=1}^{[q/2]}U^{(j)}(rs_{k})-\frac{q}{\pi }%
\int_{0}^{\infty }dy\frac{\sin (q\pi )U^{(j)}(r\cosh y)}{\cosh (2qy)-\cos
(q\pi )}\right] ,  \label{fj1}
\end{equation}%
where%
\begin{equation}
U^{(j)}(y)=4m^{D+2}y^{2}\int_{0}^{\infty }dx\,xe^{-x}f_{D/2+1}(m\sqrt{\beta
_{j}^{2}x^{2}+4y^{2}}).  \label{UjS}
\end{equation}%
The shear force vanishes for Dirichlet and Neumann boundary conditions. At
large distances from the string, $mr\gg 1$, the dominant contribution to (%
\ref{fj1}) comes from the $k=1$ term and to the leading order%
\begin{equation}
f_{j}^{(1)}\approx \frac{2\beta _{j}\left( 1-4\xi \right) }{(2\pi
)^{(D-1)/2}r}\frac{m^{D}e^{-2mr\sin (\pi /q)}}{[2mr\sin (\pi /q)]^{(D-1)/2}}.
\label{fj1large}
\end{equation}%
Note that the dependence of the shear force on the curvature coupling
parameter $\xi $ appears in the form of the coefficient $1-4\xi $. Near the
string, $mr\ll 1$, the shear stress (\ref{fj1}) behaves as $r^{1-D}$ for $%
r\ll m^{-1},|\beta _{j}|$, and as $r^{-D-1}$ for $|\beta _{j}|\ll r\ll
m^{-1} $. The divergence on the string of the self-shear stress is a
consequence of the idealized model of the cosmic string with zero thickness
core. In more realistic models, the behavior of the shear stress near the
string depends on the core structure.

\begin{figure}[tbph]
\begin{center}
\begin{tabular}{cc}
\epsfig{figure=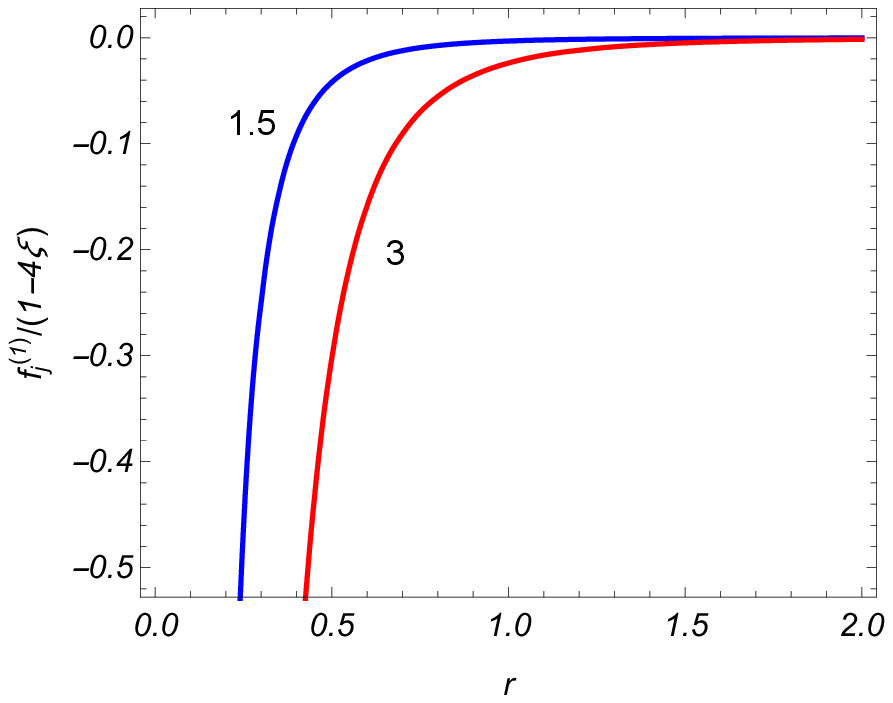,width=7.cm,height=5.5cm} & \quad %
\epsfig{figure=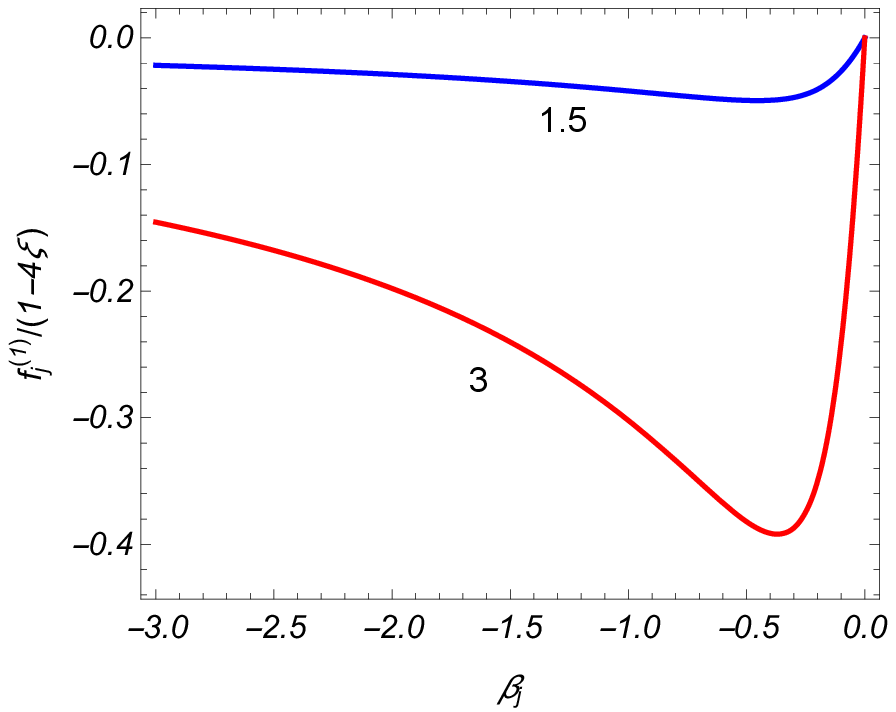,width=7.cm,height=5.5cm}%
\end{tabular}%
\end{center}
\caption{The shear stress on a single plate for a $D=4$ massless scalar
field as a function of the radial coordinate and of the Robin coefficient
(arbitrary units). The numbers near the curves correspond to the values of $%
q $. }
\label{fig3}
\end{figure}

For a massless field we have%
\begin{equation}
U^{(j)}(y)=2^{D/2+2}y^{2}\int_{0}^{\infty }dx\,\frac{\Gamma (D/2+1)xe^{-x}}{%
(\beta _{j}^{2}x^{2}+4y^{2})^{D/2+1}}.  \label{UjSm0}
\end{equation}%
In this case the decay of the shear stress at large distances is as power
law:%
\begin{equation}
f_{j}^{(1)}\approx \frac{4\beta _{j}\left( 1-4\xi \right) }{(4\pi
)^{D/2}r^{D+1}}\Gamma (D/2+1)g_{D}(q),  \label{fj1largem0}
\end{equation}%
with the function $g_{D}(q)$ defined in (\ref{cnq}). In figure \ref{fig3} we
have plotted the shear stress $f_{j}^{(1)}$ for a $D=4$ massless field as a
function of the radial coordinate (left panel, arbitrary units) and of the
coefficient in the Robin boundary condition (right panel, arbitrary units).
The left panel is plotted for $\beta _{j}=-1$ and the right panel is plotted
for $r=0.5$. The numbers near the curves are the corresponding values of the
parameter $q$.

\begin{figure}[tbph]
\begin{center}
\begin{tabular}{cc}
\epsfig{figure=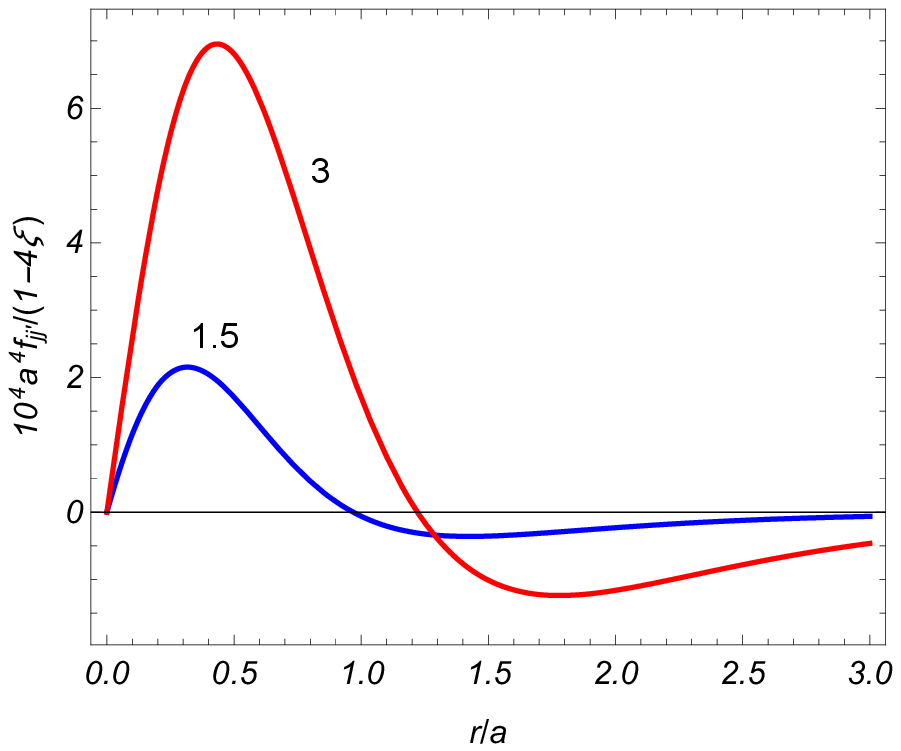,width=7.cm,height=5.5cm} & \quad %
\epsfig{figure=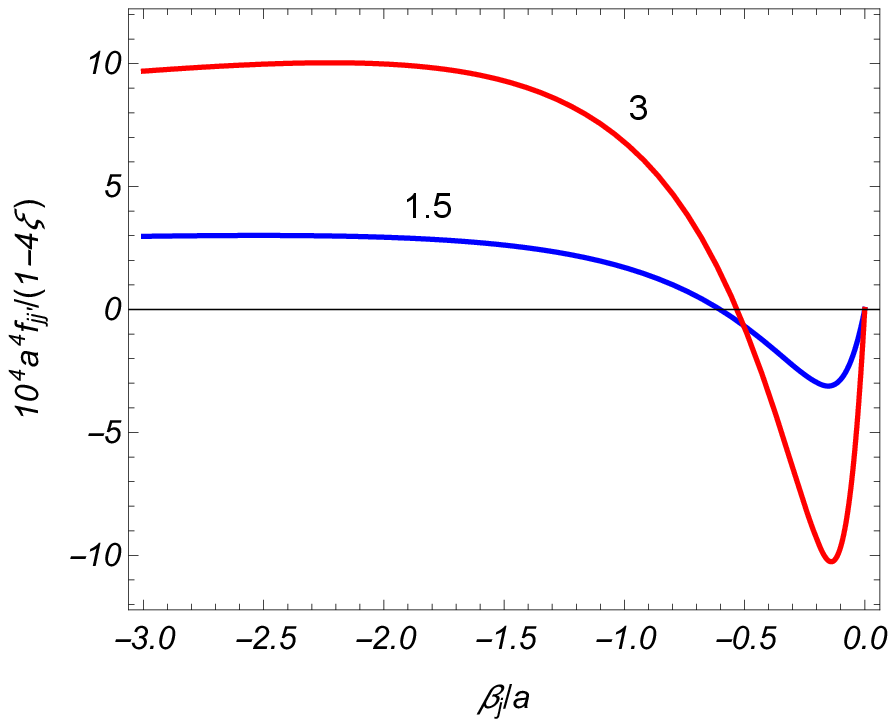,width=7.cm,height=5.5cm}%
\end{tabular}%
\end{center}
\caption{The interaction part of the shear stress in the geometry of two
plates for a $D=4$ massless scalar field as a function of $r/a$ and $\protect%
\beta _{j}/a$. The graphs are plotted for $\protect\beta _{1}=\protect\beta %
_{2}$. }
\label{fig4}
\end{figure}

In the geometry of two plates the shear force per unit surface of the plate
at $z=a_{j}$ is presented as%
\begin{equation}
f_{j}=f_{j}^{(1)}+f_{jj^{\prime }},  \label{fj}
\end{equation}%
where $f_{jj^{\prime }}=\langle T_{3}^{1}\rangle _{jj^{\prime
}}|_{z=a_{j}+0}^{z=a_{j}-0}$ is the shear stress induced by the plate at $%
z=z_{j^{\prime }}$. Note that $f_{jj^{\prime }}=0$ in the regions $%
z\leqslant z_{1}$ and $z\geqslant z_{2}$ and the force acts on the sides $%
z=a_{1}+0$ and $z=a_{2}-0$ only. By taking into account (\ref{Tmujj}) and (%
\ref{U31jj}), for the shear stress induced by the second plate one obtains%
\begin{equation}
f_{jj^{\prime }}=\frac{4\beta _{j}\left( 1-4\xi \right) }{(2\pi )^{D/2}r}%
\left[ \sideset{}{'}{\sum}_{k=1}^{[q/2]}U^{(jj^{\prime })}(rs_{k})-\frac{q}{%
\pi }\int_{0}^{\infty }dy\frac{\sin (q\pi )U^{(jj^{\prime })}(r\cosh y)}{%
\cosh (2qy)-\cos (q\pi )}\right] ,  \label{fjj}
\end{equation}%
with the function%
\begin{equation}
U^{(jj^{\prime })}(y)=\sqrt{2\pi }y^{2}\int_{m}^{\infty }du\frac{%
u^{2}(u^{2}-m^{2})^{(D-1)/2}}{c_{1}(au)c_{2}(au)e^{2au}-1}\frac{%
Z_{(D-1)/2}(2y\sqrt{u^{2}-m^{2}})}{1-\beta _{j}^{2}u^{2}}.  \label{Ujj}
\end{equation}%
Note that, unlike the self-shear, the interaction part (\ref{fjj}) is finite
everywhere, including on the string. Figure \ref{fig4} presents the
interaction part of the shear stress as a function of the ratios $r/a$ (left
panel) and $\beta _{j}/a$ (right panel) for a $D=4$ massless scalar field
and for $q=1.5,3$ (numbers near the curves) in the model with $\beta
_{1}=\beta _{2}$. The left panel is plotted for $\beta _{j}/a=-1$ and the
right panel is plotted for $r/a=0.5$.

\section{Summary}

\label{Conc}

We have investigated the effects of nontrivial topology due to a straight
cosmic string on the local characteristics of the scalar vacuum and on the
Casimir forces in the geometry of two parallel plates perpendicular to the
axis of the string. On the plates, the Robin boundary conditions are imposed
with coefficients that, in general, can differ for separate plates. In the
problem under consideration all the properties of the quantum vacuum are
encoded in two-point functions. As the first step for the evaluation of
these functions we have constructed the heat-kernel as a mode-sum over the
complete set of scalar modes. The mode-sum contains summation over the
eigenvalues for the component of the momentum perpendicular to the plates.
Unlike to special cases of Dirichlet and Neumann boundary conditions, in the
region between the plates these eigenvalues are given implicitly, as
solutions of the transcendental equation (\ref{EigEq2}). For the summation
of the corresponding series, we have employed a variant of the generalized
Abel-Plana formula that allowed us to extract explicitly the boundary-free
contribution in the Green function (see (\ref{Gdecb})) and to present the
boundary-induced part, given by (\ref{Gbn}), in terms of integrals strongly
convergent for points away from the plates. The boundary-induced
contribution to the Green function can be further decomposed into the single
boundary and second boundary-induced terms (see (\ref{Gdec2})), separately
given by (\ref{Galf}). For points away from the plates, the local geometry
in the problem at hand is Minkwoskian and, as a consequence, the
renormalization of the VEVs in the coincidence limit is reduced to the
subtraction of the corresponding VEVs in boundary-free Minkowski spacetime.
In the representations of the Green function we have provided, the
Minkowskian part is presented by the $k=0$ term in the formula (\ref{Galf})
for the boundary-free function $G_{0}(x,x^{\prime })$. Hence, the
renormalized VEVs are obtained omitting this term and taking the coincidence
limit of the arguments.

Our consideration of the VEVs of the field squared and of the
energy-momentum tensor we have started with the presence of a single plate.
The boundary-induced contributions are given by (\ref{phi2jn}) and (\ref%
{Tmuj1}). The $k=0$ terms in these expressions correspond to the VEVs for a
plate in Minkowski bulk. The remaining parts are contributions induced by
the nontrivial topology of the cosmic string. In special cases of Dirichlet
and Neumann boundary conditions the results for the geometry of a single
plate reduce to those previously derived in \cite{Beze11}. For points
outside the plates and close to the string, the VEVs are dominated by the
boundary-free contributions. The latter diverge on the string as $1/r^{D-2}$
for the field squared and like $1/r^{D}$ for the diagonal components of the
energy-momentum tensor. The boundary-induced parts in the VEVs are finite on
the string. For the field squared one has a simple relation (\ref{phi2jr0b})
with the corresponding VEV\ in Minkowski bulk. For the diagonal components
of the energy-momentum tensor in the case of a single plate the
corresponding relation is more complicated and is given by (\ref{Tmur0}).
The off-diagonal component linearly vanishes on the string. For points
outside from the string the divergences of the VEVs on the boundary coincide
with those for a plate in Minkowski bulk and the topological parts are
finite on the boundary.

The VEVs in the region between the plates are decomposed into the
boundary-free and boundary-induced contributions. The latter is given by (%
\ref{phi2b}),(\ref{U}) for the VEV of the field squared and and by (\ref%
{Tmub}),(\ref{Umu}) for the energy-momentum tensor. The $k=0$ terms in these
expressions are the corresponding VEVs in the region between two plates on
the Minkowski bulk. For points away from the boundaries, the off-diagonal
component of the energy-momentum tensor vanishes on the string. The
functions (\ref{U}) and (\ref{Umu}) in the expressions for the VEVs are
further simplified\ to (\ref{UDN}) and (\ref{UmuDNm}) in the special cases
of the Dirichlet and Neumann boundary conditions. For the off-diagonal
component of the energy-momentum tensor the corresponding function is
presented as (\ref{U31DNb}). The off-diagonal component vanishes on the
plates for Dirichlet and Neumann boundary conditions. Alternative
decompositions\ with the extracted single plate and the second plate-induced
parts in the region between the plates are given by (\ref{phi22pl}) and (\ref%
{Tmualt}).

In Section \ref{sec:Forces} we have investigated the Casimir forces acting
on the plates. Due to the nonzero off-diagonal component of the vacuum
energy-momentum tensor, in addition to the normal component, these forces
have nonzero component parallel to the boundary (shear force). The vacuum
effective pressure on the plates, corresponding to the normal component of
the Casimir force, is given by (\ref{Pj}) with the function (\ref{Fj}). The
corresponding forces are attractive for $P_{j}<0$ and repulsive for $P_{j}>0$%
. Unlike the problem on the Minkowski bulk, the forces acting on the
separate plates, in general, do not coincide if the corresponding Robin
coefficients are different. Another difference is that in the presence of
the cosmic string the Casimir forces for Dirichlet and Neumann boundary
conditions differ. In these special cases the functions in the expression
for the pressures are simplified to (\ref{FjD}) and (\ref{FjN}). For a
massless field the corresponding formulas take the form (\ref{FjDm0}) and (%
\ref{FjNm0}). For odd values of the spatial dimension the corresponding
series are expressed in terms of elementary functions (see (\ref{FjDN}) for $%
D=4$). For Dirichlet boundary condition the Casimir pressure do not depend
on the curvature coupling parameter. This is not the case for Neumann
boundary condition. A new qualitative feature induced by the cosmic string
is the appearance of the shear stress acting on the plates. The
corresponding force is directed along the radial coordinate. The shear force
vanishes for Dirichlet and Neumann boundary conditions. In the geometry of a
single plate, the corresponding stress is given by (\ref{fj1}) with the
function (\ref{UjS}). At large distances from the string the shear force
decays as $e^{-2mr\sin (\pi /q)}$ for a massive field and like $1/r^{D+1}$
for a massless field. In the geometry of two plates the shear stress is
decomposed into two contributions (see (\ref{fj})). The first one
corresponds to the self-stress and the second one is induced by the presence
of the second plate. The latter is given by (\ref{fjj}) and (\ref{Ujj}).
Depending on the parameters of the problem, the radial component of the
shear force can be either positive or negative.

The regularization procedure we have used is based on the point-splitting.
An alternative regularization procedure, widely discussed in the literature,
uses the zeta function for the evaluation of global quantities, like the
total vacuum energy, or the local zeta function for the investigation of the
local VEVs (for example, the energy density and stresses). In appendix \ref%
{sec:Zeta}, on the base of the heat kernel, we have evaluated the
off-diagonal and local zeta functions in the problem under consideration.

\section*{Acknowledgment}

E.R.B.M. thanks Conselho Nacional de Desenvolvimento Cient\'{\i}fico e
Tecnol \'{o}gico (CNPq) for partial financial support. A.A.S. was supported
by the State Committee of Science Ministry of Education and Science RA,
within the frame of Grant No. SCS 15T-1C110.

\appendix

\section{Local zeta function}

\label{sec:Zeta}

By using the heat kernel $K(x,x^{\prime };s)$ from (\ref{heat3}), we can
evaluate a more general object, the off-diagonal\ zeta function (see, for
example, \cite{More97})
\begin{equation}
\zeta (x,x^{\prime };s)=\frac{\mu ^{2s}}{\Gamma \left( s\right) }%
\int_{0}^{\infty }du\,u^{s-1}K(x,x^{\prime };u),  \label{zetaK}
\end{equation}%
where $\mu $ is a mass parameter introduced by dimensional reasons. By
taking into account the expression (\ref{heat3}) for the heat kernel, one
gets%
\begin{equation}
\zeta (x,x^{\prime };s)=\frac{\left( 2\pi \right) ^{\frac{1-D}{2}}\mu ^{2s}}{%
2^{s-2}\Gamma \left( s\right) }\left[ \sum_{k}S(w_{k},x,x^{\prime };s)-\frac{%
q}{2\pi }\sum_{l=\pm 1}\int_{0}^{\infty }dy\frac{\sin (q\pi +lq\Delta
\varphi )S(w_{y},x,x^{\prime };s)}{\cosh (qy)-\cos (q\pi +lq\Delta \varphi )}%
\right] ,  \label{zeta}
\end{equation}%
with the function%
\begin{equation}
S(w,x,x^{\prime };s)=\frac{1}{2a}\sum_{p=1}^{\infty }\frac{%
(m^{2}+k_{z}^{2})^{\frac{D-1}{2}-s}f_{\frac{D-1}{2}-s}(\sigma (w)\sqrt{%
m^{2}+k_{z}^{2}})}{1+\cos [y_{p}+2\tilde{\gamma}_{j}(y_{p})]\sin
(y_{p})/y_{p}}g_{j}(z,z^{\prime },k_{z}),  \label{Szeta}
\end{equation}%
and $k_{z}=y_{p}/a$. Note that $G(x,x^{\prime })=\zeta (x,x^{\prime };1)/\mu
^{2}$.

Further transformation of the function is similar to that we have used for
the function $S(w,x,x^{\prime })$ in (\ref{Szz}). By using the summation
formula (\ref{sumfor}), the following decomposition is obtained
\begin{equation}
S(w,x,x^{\prime };s)=S_{0}(w,x,x^{\prime };s)+S_{j}(w,x,x^{\prime
};s)+S_{jj^{\prime }}(w,x,x^{\prime };s),  \label{S0z}
\end{equation}%
where the expressions for the functions in the right-hand side are obtained
from the corresponding expressions in (\ref{S0}) by the replacement $%
(D-3)/2\rightarrow (D-1)/2-s$. As a consequence, the off-diagonal zeta
function in the region between the boundaries is presented as%
\begin{equation}
\zeta (x,x^{\prime };s)=\zeta _{0}(x,x^{\prime };s)+\zeta _{\mathrm{b}%
}(x,x^{\prime };s),  \label{zetadec}
\end{equation}%
where $\zeta _{0}(x,x^{\prime };s)$ is the corresponding function for the
geometry in the absence of boundaries and the contribution $\zeta _{\mathrm{b%
}}(x,x^{\prime };s)$ is induced by the boundaries. The expressions for the
functions $\zeta _{0}(x,x^{\prime };s)$ and $\zeta _{\mathrm{b}}(x,x^{\prime
};s)$ are obtained from (\ref{zeta}) by the replacements $%
S(w_{k},x,x^{\prime };s)\rightarrow S_{0}(w_{k},x,x^{\prime };s)$ and $%
S(w_{k},x,x^{\prime };s)\rightarrow S_{\mathrm{b}}(w_{k},x,x^{\prime };s)$,
respectively, with
\begin{equation}
S_{\mathrm{b}}(w,x,x^{\prime };s)=S_{j}(w,x,x^{\prime };s)+S_{jj^{\prime
}}(w,x,x^{\prime };s).  \label{Sbz}
\end{equation}%
The integral representation for the latter is obtained from (\ref{Sb}) by
the replacement $(D-3)/2\rightarrow (D-1)/2-s$.

From (\ref{zetadec}) for the local zeta function $\zeta (x;s)=\zeta (x,x;s)$
one gets%
\begin{equation}
\zeta (x;s)=\zeta _{0}(x;s)+\zeta _{\mathrm{b}}(x;s),  \label{zetadec2}
\end{equation}%
where%
\begin{eqnarray}
\zeta _{0}(x;s) &=&\zeta _{\mathrm{M}}(x;s)+\frac{2^{2-s}\mu ^{2s}m^{D-2s}}{%
\left( 2\pi \right) ^{D/2}\Gamma \left( s\right) }\left[ \sideset{}{'}{\sum}%
_{k=1}^{[q/2]}f_{D/2-s}(2mrs_{k})\right.  \notag \\
&&\left. -\frac{q}{\pi }\int_{0}^{\infty }dy\frac{\sin (q\pi
)f_{D/2-s}(2mr\cosh y)}{\cosh (2qy)-\cos (q\pi )}\right] ,  \label{zeta0}
\end{eqnarray}%
is the local zeta function in the absence of boundaries and $\zeta _{\mathrm{%
M}}(x;s)$ is the corresponding function in Minkowski spacetime. The second
term in the right-hand side of (\ref{zeta0}) is induced by the nontrivial
topology of the cosmic string. In the region between the boundaries the
boundary-induced contribution in (\ref{zetadec2}) is given by
\begin{equation}
\zeta _{\mathrm{b}}(x;s)=\frac{\left( 2\pi \right) ^{\frac{1-D}{2}}\mu ^{2s}%
}{2^{s-3}\Gamma \left( s\right) }\left[ \sideset{}{'}{\sum}_{k=0}^{[q/2]}S_{%
\mathrm{b}}(s_{k},x;s)-\frac{q}{\pi }\int_{0}^{\infty }dy\frac{\sin (q\pi
)S_{\mathrm{b}}(\cosh y,x;s)}{\cosh (2qy)-\cos (q\pi )}\right] ,
\label{zetalf}
\end{equation}%
with the function%
\begin{eqnarray}
S_{\mathrm{b}}(y,x;s) &=&\frac{1}{4}\int_{m}^{\infty }du\frac{%
(u^{2}-m^{2})^{(D-1)/2-s}}{c_{1}(au)c_{2}(au)e^{2au}-1}Z_{\frac{D-1}{2}%
-s}(2ry\sqrt{u^{2}-m^{2}})  \notag \\
&&\times \left[ 2+\sum_{j=1,2}e^{2u|z-a_{j}|}c_{j}(au)\right] .  \label{Sbz2}
\end{eqnarray}%
For points away from the boundaries this contribution is finite at $s=1$ and
the renormalization of the local VEVs is reduced to the one in the
boundary-free geometry. In particular, for the VEV\ of the field squared one
gets
\begin{equation*}
\langle \phi ^{2}\rangle =\lim_{s\rightarrow 1}\mu ^{-2}\left[ \zeta
(x;s)-\zeta _{\mathrm{M}}(x;s)\right] .
\end{equation*}%
With the boundary-free zeta function (\ref{zeta0}), this leads to the result
(\ref{phi2cs}) for the boundary-free geometry and to the result (\ref{phi2b}%
) in the region between two plates. For the evaluation of the VEV of the
energy-momentum tensor, in addition to the local zeta function $\zeta (x;s)$%
, one needs the off-diagonal zeta function $\zeta (x,x^{\prime };s)$. The
latter is required for the evaluation of the first term in the right-hand
side of (\ref{Tmuj}). It is presented as $\lim_{s\rightarrow
1}\lim_{x^{\prime }\rightarrow x}\partial _{\mu ^{\prime }}\partial _{\nu
}\zeta (x,x^{\prime };s)$.

The renormalization procedure for the boundary-free cosmic string geometry
within the framework of the zeta function approach and the comparison with
the point-splitting scheme have been discussed in \cite{Zerb96}. Note that
the expression for the local zeta function in the boundary-free cosmic
string geometry, $\zeta _{0}(x;s)$, given in the second paper of Ref. \cite%
{Zerb96}, presents this function in the form of the series over the product $%
mr$. The local zeta function in a related geometry of a wedge with
reflecting boundaries is considered in \cite{Nest02}. Note that, unlike to
the case of the cosmic string geometry, the problem with the wedge is not
homogeneous along the azimuthal direction.

\end{document}